\documentclass[11pt,a4paper]{article}
\usepackage{jheppub} 
\usepackage{graphicx}  
\usepackage{dcolumn}   
\usepackage{bm}        
\usepackage{amssymb}   
\usepackage{amsmath}
\usepackage{xspace}
\usepackage{epstopdf}
\usepackage{wrapfig}
\usepackage{color}
\usepackage{bbm}
\usepackage{float}
\usepackage{appendix}
\usepackage{braket}
\usepackage{subfigure}

\usepackage{times}
\usepackage{color}

\newcommand{\gtilde}{$\tilde{g}$}
\newcommand{\s}{$S$}
\newcommand{\sbar}{$\bar{S}$}
\newcommand{\sdm}{$S$DM}
\newcommand{\ssbar}{$S,\,\bar{S}$}

\newcommand{\be}{\begin{equation}}
\newcommand{\ee}{\end{equation}}

\begin{document} 

\title{A Stable Sexaquark: \\Overview and Discovery Strategies}

\author
{Glennys R. Farrar }
\affiliation{Center for Cosmology and Particle Physics,
Department of Physics\\
New York University, NY, NY 10003, USA}

 \emailAdd{gf25@nyu.edu}
 \arxivnumber{}

\abstract{
A compact, deeply bound state of $uuddss$ quarks would, if it exists, be stable or very long-lived.  
Lattice QCD calculations cannot yet determine whether this neutral, spin-0, flavor singlet di-baryon (``S") exists, and empirical models give highly divergent predictions for the mass of a possible compact configuration distinct from a $\Lambda$-$\Lambda$ molecule.
Surprisingly, experimental observations to date would not have detected it.  The strongest present laboratory constraint comes from the lower bound on the H-dibaryon formation time in a doubly strange hypernucleus, but this is insufficient to discover or exclude a compact, long-lived \s.  In this paper we develop several experimental strategies to discover or constrain such a state, some of which are more generally applicable to any new long-lived neutral hadron.  If a stable \s\ exists, it could comprise the dark matter as well as potentially resolve the muon g-2 anomaly.
} 

\maketitle 

\section{Introduction}
\label{sec:intro}

Elementary particles consisting of four or more quarks and antiquarks
have been of considerable theoretical and experimental interest in
recent years. 
In 2003 Belle discovered an extremely narrow four-quark state,  
$X(3872)$, with quark content $c \bar c u \bar u$~\cite{BelleTetraquark03}.
This discovery has been confirmed by many experiments.\footnote{For a detailed review on $X(3872)$, also known as
${{\chi}_{{c1}}{(3872)}}$ see the PDG\kern0.1em{Live}
\url{
https://pdglive.lbl.gov/Particle.action?init=0&node=M176&home=MXXX025}.}
In 2015 LHCb discovered a narrow pentaquark state with quark content 
$c \bar c u u d$~\cite{LHCbPentaquark15}. In 2019 using significantly larger
statistics, LHCb resolved the data into three narrower states with the same
quark content~\cite{LHCbPentaquark19}.
Many other multiquark states have been identified in the years 2003-2017~\cite{karliner+review18}. They all involve a heavy quark and antiquark,
$c \bar c$ or $b \bar b$. Most of these states are likely to be
hadronic molecules, i.e. heavy-quark analogs of the deuteron, 
with two color-singlet hadrons attracting each other via light
meson exchange.

The situation changed dramatically with the recent discovery by LHCb
of the $T_{cc}$ tetraquark with quark content $c c \bar u \bar d$~\cite{LHCb:2021vvq,LHCb:2021auc}.
This discovery provides significant support to the theoretical 
consensus that the yet-to-be-observed analogous $b b \bar u \bar d$ state is 
 a deeply bound compact tetraquark, as opposed to a hadronic molecule.
This state is expected to lie well below the $B B$ threshold, 
making it stable against strong decay 
~\cite{KarlinerRosDoubChBar17,EichtenQuigg17,Czar+Tetrons17,Hudspith+LatExotic20,Lesk+LatDoubBot19}.
It is natural to ask if any deeply-bound, compact, multi-quark states involving only light $u$, $d$, $s$ quarks may be possible.   Unlike for hadrons containing heavy quarks, theory for multi-light-quark states is not presently sufficiently accurate to make reliable predictions so experiment must be the guide.  

This paper is devoted to the experimental detection of a stable bound state  of six  light quarks, $u_\uparrow u_\downarrow d_\uparrow d_\downarrow s_\uparrow s_\downarrow $, having a mass low enough to be absolutely stable, or effectively stable with a lifetime greater than the age of the Universe, or simply a lifetime so long that it would not be detected in searches for neutral decaying particles.\footnote{By contrast, in the context of the tetraquark and pentaquark states, ``stable" refers to decaying via an electromagnetic or weak rather than strong interaction.} The postulated stable $uuddss$ state is designated \s\ or ``sexaquark"~\cite{fS17}.\footnote{``Hexaquark" is the generic term for any dibaryon or triple-$q \bar{q}$ hadron.  The sexaquark discussed here is a specific example of a hexaquark.  The name sexaquark, ``$S$", was chosen for the stable state considered here because its phenomenology is entirely different from other hexaquarks; using the symbol H for this particle would cause confusion with the H-dibaryon with lifetime $\approx 10^{-10}$ s or generic hexaquarks, as well as with the Higgs.  The word sexaquark is appropriate because it evokes six, stable, singlet, scalar, strange and strong -- all attributes of the state.  An additional benefit of using the Latinate ``sexa" is that then the Greek prefix series tetra-, penta-, hexa-  remains used for states consisting of $q \bar{q}$ or $qqq$ plus $q \bar{q}$ pairs. }  
The stability of an \s\ can result from its mass being too low for baryon-number-conserving decay, or from its decay requiring a doubly-weak interaction and/or the transition amplitude being suppressed due to small spatial overlap or tunneling barrier~\cite{fzNuc03}.   Note that both an \s\ and a di$-\Lambda$ molecule may exist.  

This paper has two key parts:  1) establishing that a hadron with properties of the sexaquark would not have been detected by existing experiments and 2) proposing several experimental strategies to rectify this deficiency in our experimental knowledge.
The general experimental challenge to discovering an \s\ or excluding its existence is that \s's are neutral and similar in mass to neutrons, but interact less and are much less abundant.   \s's do not call enough attention to themselves in high energy experiments to have been recognized as a new particle among the plethora of final particles.  As we shall discuss in detail, dedicated searches for the H-dibaryon (a hypothesized particle~\cite{jaffe:H} having the same quantum numbers as the \s\ but with a weak-interaction lifetime) either employed signatures not applicable to the stable \s\ or had insufficient sensitivity.  

The experimental search for a sexaquark has motivations beyond the importance of having a correct picture of the stable hadrons.  We recently showed~\cite{fg-2_22} that the muon g-2 anomaly~\cite{g-2PRL21} can be due to missed hadronic final states in $e^+e^-$ annihilation resulting in a too-low value of the hadronic vacuum polarization (HVP) as determined by the R-ratio measurements~\cite{Aoyama+20}.   This missed-hadronic states mechanism also explains why the latest lattice QCD calculations of the HVP~\cite{BMWHVP21,Alexandrou+ETMC22} agree with the g-2 measurement but not with the direct R-ratio value.  In order for hadronic final states to be missed, a significant portion of the final energy needs to be carried by neutral long-lived hadrons~\cite{fg-2_22}.  The sexaquark is the only concrete candidate so far for filling the requirements of that role, further motivating the experimental searches proposed here.
Moreover a neutral stable sexaquark potentially provides dark matter without physics beyond the Standard Model~\cite{fHDM02}:  it can reasonably have the observed relic abundance~\cite{fDMtoB18} and is consistent with existing limits on dark matter interactions~\cite{fwx_SDM20,xf21a,xf21b}.  

The organization of this paper is as follows.  We begin with several short sections reviewing information needed for thinking about laboratory sexaquark detection, so that the paper is self-contained.  Section~\ref{sec:overview} provides general background and context.
 In Sec.~\ref{sec:mass} we briefly survey calculations of the mass of the H-dibaryon and the \s.  In Sec.~\ref{sec:prop-xcn} we discuss the properties of the \s\ that account for its elusiveness and are pertinent to its experimental detection, and estimate production and scattering cross sections.  Section~\ref{sec:gtilde} discusses the amplitude for interconversion between \s\ and two baryons, which determines the cross section for  \s\ production in low energy experiments and the formation time of \s\ in a hypernucleus; this interconversion amplitude is also relevant for whether sexaquark dark matter can survive in the hot hadronic phase of the Early Universe.  Section~\ref{sec:astro} summarizes the principle constraints on the scattering cross section from cosmology, astrophysics, and dark matter (DM) direct detection experiments which apply if DM is composed predominantly of sexaquarks.  These preliminaries lay the groundwork for Sec.~\ref{sec:expts}, the central content of the paper, in which strategies to discover the \s\ in laboratory experiments are proposed.  Section~\ref{sec:sum} summarizes the paper.  Details of certain lengthy calculations are given in the Appendices.

\vspace{-0.04in} 
\section{Overview}
\label{sec:overview}
\vspace{-0.04in}  

The proposed \s\  is a spin-0, flavor singlet, parity-even boson with charge Q=0, baryon number B=2, and strangeness S=-2.  Unlike the tetraquark, pentaquark and the heptaquark states, the sexaquark is composed of valence quarks and no valence antiquarks.   Analog states of the \s\ with one or more of the light ($uds$) quarks substituted by a $c$ or $b$ of the same charge, would have a short lifetime  ($\lesssim 10^{-8}$ s) since direct decay at the quark level such as $c \rightarrow d $ or $b\rightarrow c,u$ is not kinematically suppressed.   Moreover since there are 18 different color, flavor, spin combinations using $u,d,s$ quarks, totally antisymmetric color singlet states with baryon number up to 6 could be potentially interesting.   The phenomenology of these states is an entirely different topic and is not the subject of the present paper.

A sufficiently stable \s\  is potentially a good dark matter candidate \cite{fDMtoB18,fSDM_ICRC17,fSDM22}, a possibility we first noted in \cite{fArkady03}.  The ratio of densities of \s\ dark matter (\sdm) and baryons in the Universe   
can be estimated from statistical equilibrium in the quark-gluon plasma using known parameters from QCD~\cite{fDMtoB18} and is consistent with observation.  Primordial nucleosynthesis limits on unseen baryons are satisfied as long as ambient \s's do not participate actively in nucleosynthesis, i.e., \s\ does not form bound states with light nuclei D, T or He. These and related topics are briefly summarized in Sec.~\ref{sec:astro}.

Whether the \s\ is stable enough to be dark matter depends on its mass and its dissociation amplitude to two baryons, \gtilde;  as discussed in Sec.~\ref{sec:gtilde}, theoretical estimates suggest \gtilde\ $\lesssim 10^{-6}$.  An \s\ lighter than deuterium cannot decay without violating baryon number conservation and so would be absolutely stable.   Even if the \s\ is not absolutely stable, it can be the dark matter or at least a component of it, if its lifetime is long enough compared to the age of the Universe.  This is assured if $m_S < m_p+m_e + m_\Lambda = 2054.46$ MeV, the lightest mass compatible with a $\Delta$S = 1 decay, below which the decay rate $\sim G_F^4 \tilde{g}^2 $, causing the \s\ to be effectively stable~\cite{fzNuc03}.  Two separated baryons with the same quark content as the \s\ must have a mass $\geq  2 m_\Lambda = 2231.36$ MeV, so if the 6 quarks are bound by $\gtrsim 176.9$ MeV, the state is effectively stable.\footnote{Lifetime estimates taking into account various processes and considering a range of masses were derived in~\cite{GDHprd86,DGHCyg86}, however the \gtilde$^2$ suppression of the decay rate was not included so the lifetimes obtained should be multiplied by $\approx 5/ \tilde{g}^{2}$, where the factor 5 reflects that $\sqrt{4/5}$ of the \s\ wavefunction consists of colored-octet baryons which are not available final states~\cite{}.}

Deep binding is facilitated by the unique symmetry structure of the \s.  Fermi statistics give the 6-quark combination $uuddss$ a privileged status.  Uniquely among 6-light-quark states, the spatial wavefunction of the \s\ can be totally symmetric at the same time that the color, flavor, and spin wave functions are simultaneously individually totally antisymmetric.  (Other color-singlet 6-light-quark states are not spatially symmetric, e.g., the deuteron is a loosely bound pair of nucleons, and states containing still heavier quarks are not stable to weak decay and hence would not be of interest as dark matter candidates.)   When mass-splitting is governed by hyperfine interactions (at the heart of general most-attractive-channel arguments, e.g.,~\cite{RDSTumbling80,PeskinVacAlign80,PreskillMAC81}, and fundamental in explaining the baryon mass hierarchy in QCD~\cite{DGG75}) the most deeply-bound state of a system of fermions is the one in the smallest allowed representation of the color, spin and internal symmetries.    Thus the \s, being a singlet in both flavor and spin, should be the most-tightly bound 6-quark state, analogous to the deep binding of the $^4$He nucleus compared to two deuterons, and to the singlet state in positronium being lower in energy than the triplet. 

The tendency for strong attraction in the $u_\uparrow u_\downarrow d_\uparrow d_\downarrow s_\uparrow s_\downarrow $ system was pointed out by R. Jaffe 40 years ago.  Using the bag model and a one-gluon-exchange calculation of QCD hyperfine splitting~\cite{jaffe:H}, he estimated the mass of the state, which he called the H-dibaryon, to be $\approx 2150$ MeV.  With this mass, the H can decay via a singly-weak interaction and was estimated to have a typical weak interaction lifetime $\sim 10^{-10}-10^{-8}$ s~\cite{GDHprd86}.  

Numerous experiments to discover the H-dibaryon gave null results.   Observation of hyperon decay products from an $\rm{S}\!=\! -2$ hypernucleus~\cite{ahn+hypernucBNL01,takahashi+hypernucKEK01} is evidence against the existence of an H-dibaryon with a formation time that is shorter than the hyperon lifetime.      
The limit on production of a narrow $\Lambda p \pi^-$ resonance in the final states of $\Upsilon$ decay \cite{BelleH13}, is  evidence against an H-dibaryon with  $m_S > m_\Lambda + m_p + m_\pi = 2193.5$ MeV.   Searches for a new neutral decaying particle \cite{NA3longLiveNeuts86,bernstein+LongLivedNeut88,belz+96,KTeVHdecay99} failed to find an H-dibaryon with a typical weak interaction lifetime and mass in the expected range.  A beam dump plus time-of-flight experiment to search for new long-lived neutral particles \cite{gustafson} might have had the sensitivity to discover an H-dibaryon or \s, except that that experiment explicitly excluded consideration of masses less than 2 GeV, to mitigate the overwhelming neutron background. 

In sum,  H-dibaryon searches to date exclude an H-dibaryon with a typical weak interaction lifetime, $\sim 10^{-10}-10^{-8}$ s but do not exclude the existence of a long-lived state with the same quantum numbers.  

\vspace{-0.04in} 
\section{Mass and stability of the sexaquark}
\label{sec:mass}
\vspace{-0.04in}  

Eventually, lattice QCD should be able to answer the question of whether a stable sexaquark exists.  The challenges in applying lattice QCD are keeping the noise at an acceptable level while reaching the infinite volume and continuum limits with realistic quark masses.   The problems grow rapidly with the number of constituents~\cite{LepageTASI}, and lighter quarks increase the noise.  The best conventional lattice gauge calculation to date, that of the NPLQCD collaboration~\cite{Beane+13} using the L\"uscher method, had a pion mass $\sim$800 MeV.   While far from physical quark masses, this calculation confirms the tendency to relatively deep binding, with the B=2, S=-2 channel~\cite{Beane+13} having a binding energy of 80 MeV relative to two $\Lambda$'s. 
Even with an 800 MeV pion, Ref.~\cite{Beane+13} has only a short plateau-like region before the noise takes over, showing the challenge of reaching the physically realistic, highly relativistic situation where chiral symmetry breaking condensates are predominantly responsible for the mass.  The HAL QCD approach~\cite{HALQCD12} fits two baryon correlation data on the lattice with an effective scattering potential.  Their calculations to date, using realistic quark masses~\cite{SasakiHALQCD20}, do not provide evidence for or against a compact, deeply bound \s\ particle in the correlation function.   Such evidence would be suppressed first by a simple Clebsch-Gordan factor of 1/40  owing to the overlap of a state of two color-singlet $\Lambda$'s with a fully antisymmetrized six quark state with \s\ quantum numbers, as well as further dynamically suppressed owing to the small spatial overlap of the wavefunctions (see discussion of $\tilde g$ in Sec.~\ref{sec:gtilde} below); methods to improve the sensitivity are under investigation.\footnote{T. Hatsuda, private communication.}  The sensitivity of studies using the  L\"uscher method, e.g.,~\cite{Beane+13}, is also reduced by the inhibition of \s\ production from a pair of S=-1 baryons, embodied in the small value of \gtilde.

Unfortunately, there is no good analog system for empirically estimating the mass  of the \s\  based on other hadron masses.  The \s\ is a scalar, so chiral symmetry breaking has no formal implications for its mass, and its mass has no \textit{a priori} relation to the masses of baryons.   Model predictions for the H-dibaryon mass cover a wide range from stable to unbound ($m_S > 2231.4$ MeV).  Reference~\cite{kochelev99} calculated a mass of 1718 MeV in the instanton liquid model, disputing the earlier claim~\cite{takeuchiOka90} that three-body repulsion due to the light-quark-instanton unbinds the H entirely.   QCD sum rule calculations have predicted masses from $\sim$1.2 GeV \cite{azizi+19} to $\sim \!2 m_\Lambda = 2230$ MeV\cite{kodama+94}.  An early constituent quark model with hyperfine interactions gave 2.18 GeV \cite{rosner86}, naive diquark phenomenology resulted in estimates ranging from $\sim1.2$ GeV to $\sim2.17$ GeV \cite{gross+strumia+18} and a recent detailed study with diquarks found a mass of 1883 MeV \cite{buccella20}.  

To summarize, some degree of binding in the flavor- and spin-singlet di-baryon channel is predicted in most models, yet such a state appears to be excluded experimentally if it decays with a typical weak interaction lifetime.   This means either that almost all model predictions are wrong, or the state is long enough lived to have escaped detection in existing searches.  Thus the experimental search for a stable or effectively stable \s\ is highly motivated; as we shall see in the following, the mass range $1.85-2.2$ GeV is most promising. 

\section{Properties and Cross Sections} 
\label{sec:prop-xcn}

Three attributes of a stable \s\ make it very difficult to detect:\\ 
$\bullet$ The \s\ is neutral and a flavor singlet, so it does not couple to photons, pions and most other mesons, nor does it leave a track in a detector. \\
$\bullet$ The \s\ has no pion cloud and may be more compact than ordinary baryons; the amplitude for interconversion between \s\ and baryons is small. \\
$\bullet$ The mass of the \s\ makes it difficult to distinguish from the much more copious neutron. 

Being a flavor-SU(3) singlet, the \s\ cannot couple to flavor-octet mesons, except through an off-diagonal coupling transforming it to a much heavier flavor octet dibaryon.  
At low momenta, as relevant for dark matter interactions, the $SN$ interaction can receive contributions from the exchange of glueballs (masses $\gtrsim 1.5$ GeV), the flavor singlet superposition of $\omega$ and $\phi$ vector mesons denoted $\phi_0$ (mass $\approx 1$ GeV) and the $f_0$ -- the very broad isosinglet scalar (also known as $\sigma$) with mass $\sim 500$ MeV which is a tetraquark or di-meson molecule.  Due to the extended nature of the $f_0$ and the fact that pions do not couple to the \s, the $f_0$'s contribution to \s\ interactions should be small compared to that of the $\phi_0$, particularly if the $f_0$ is a two pion resonance.  
Exchange of $\phi_0$, which is a vector meson,  produces a repulsive nucleon-nucleon or \s-\s\ interaction.  However the sign of the \s-$\phi_0$ coupling need not be the same as that of the $N$-$\phi_0$ coupling, so the \s-nucleon interaction can be attractive or repulsive.  There is no fundamental approach to predict the strength of the $\phi_0$ coupling to \s\ (or, for that matter, the $\phi_0$ coupling to other hadrons).  Without this, we
cannot model $b$ appearing in Eq.~\eqref{rN} below or calculate the cross section for non-relativistic $\phi_0$-mediated DM-nucleon and DM-nucleus scattering.   
 
The charge radius of a nucleon, $\sim$0.9 fm in spite of its Compton wavelength being 0.2 fm, is a measure of the cloud of pions surrounding it.  The fact that the \s\ does not couple to the octet of pseudoscalar mesons, on account of  its being a flavor singlet, implies that the \s\ is not surrounded by a pion cloud.  Therefore it would be expected to have a smaller spatial extent than octet mesons and baryons.  Naively parameterizing the radius $r_X$ of a particle $X$ as a linear superposition of its Compton wavelength $\lambda_X$ plus that of the lightest meson $M$ to which it couples strongly, 
\be
\label{rN}
r_X = \lambda_X + b \, \lambda_M~,
\ee
and fitting to the nucleon charge radius, we  find $b=0.45$.  For $m_S \approx 2$ GeV, the Compton wavelength $\lambda_S \approx 0.1$ fm, so Eq.~(\ref{rN}) with $b = 0.45$ would imply that the \s\ would have a spatial extent $r_S \approx 0.2-0.3$ fm for $\phi_0$ or $f_0$ mesons respectively.  There is no contradiction between being more compact and being more deeply bound relative to other light-quark hadrons, since the short-distance QCD attraction, $\sim \ln ({\rm r)/r}$, compensates the increased zero-point kinetic energy by virial arguments. 

Our understanding of non-perturbative QCD is presently inadequate to enable a first principles prediction of hadron cross sections at energies in and above the resonance regime.  Naively invoking a black-disk model using the charge radii fails badly for nucleon-nucleon and meson-nucleon scattering, for example predicting an elastic $pp$ cross section more than a factor 5 greater than observed at  $E_{CM} = 10$ GeV and even more discrepant for $\pi p$ and $K p$.\footnote{The black-disk model for scattering of hadrons having radii $R_1$ and $R_2$ predicts $\sigma_{\rm tot} = \pi (R_1 + R_2)^2 = 2 \sigma_{\rm el} = 2 \sigma_{\rm inel}$;  for data see \url{https://pdg.lbl.gov/2020/hadronic-xsections/}.  See~\cite{StodolskyVHExcn17} for discussion and overview of the status of modeling cross sections at very high energy where a black disk model should eventually be valid.}   It is not surprising that a black-disk model fails at low energy because hadrons are color-singlets so their couplings and scattering amplitudes are non-zero only due to interactions between color fluctuations in the respective hadrons.   With $r_S$ potentially considerably smaller than $r_p$, the probability of color fluctuations in both hadrons intersecting and producing an interaction could be quite a bit smaller than in a $pp$ collision;  in that case the sexaquark and nucleon or nucleus would be more transparent to each other and their cross section much smaller. With this in mind, in our discussion of experimental strategies we will consider sexaquark cross sections ranging from the corresponding nucleon cross section to much smaller.  As we shall see, appropriately designed experiments can find sexaquarks over a large range of scattering cross sections.  
 
In addition to being electrically neutral, the \s\ has spin-0 so its magnetic dipole moment is zero.  In the equal $u,\,d,\,s$ quark mass limit, the spatial wavefunction of the \s\ is perfectly symmetric and its charge radius vanishes. Thus its charge radius is very small and its coupling to photons is suppressed by SU(3) flavor symmetry and powers of $r_S$, the \s\ radius.  For $r_S \lesssim 0.4$ fm, $S$ coupling to $\gamma$'s can only be non-negligible for momentum-transfer $\gtrsim \mathcal{ O}$(0.5 GeV).  Thus elastic scattering of \s\ with gammas or electromagnetic fields is not astrophysically or cosmologically relevant.   Note that since the \s\ is an isospin singlet, its interactions are the same with neutrons and protons.

\section{Conversion between \s\ and baryons} 
\label{sec:gtilde}
Processes that require low-relative-momentum interconversion of \s\ and baryons (e.g., fusion or dissociation) are strongly suppressed, even though the scattering cross section of an \s\ with nucleons and nuclei can be within an order of magnitude of similar hadronic reactions.  
Such interconversion processes include: 
\\
$\bullet$ Fusion of hyperons to an \s\ in a doubly strange hypernucleus;
\\
$\bullet$ Breakup of \s\ dark matter in the hot hadronic medium after \s DM formation at the end of the quark gluon plasma era;
\\
$\bullet$ Production of \s\ in a reaction like $K^- p \rightarrow S \bar{\Lambda}$;
\\
$\bullet$ Decay of nuclei to \s\ if kinematically allowed, e.g., $d \rightarrow S + e^+ + \nu_e$;
\\
$\bullet$ Decay of the \s\ if it is not absolutely stable.

\begin{figure}[t]
     \centering
     \includegraphics[width=0.68\textwidth]{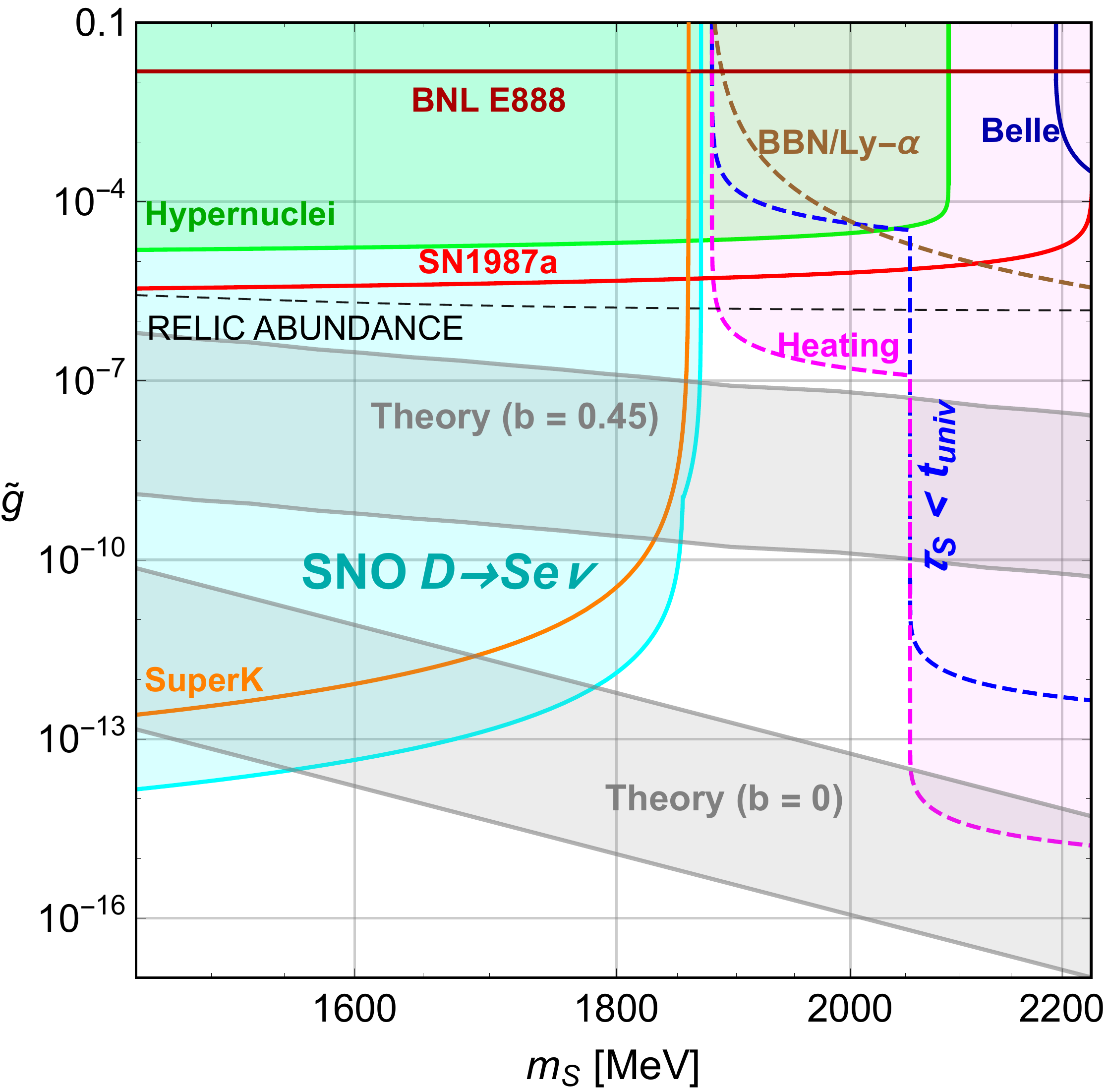}
        \caption{Excluded and predicted regions of \gtilde, reproduced from~\cite{fw22a}. The upper grey band is the theoretical prediction for \gtilde, with the band width reflecting different estimates of the tunneling suppression.  The colored regions show constraints described in the text and in Ref.~\cite{fw22a}.  Regions bounded by solid lines are excluded independently of whether the \s\ comprises DM, while regions bounded by dashed lines would be excluded if \s\ comprises 100\% of the DM. }
        \label{fig:gtilde}
\end{figure}

The interconversion between an \s\ and baryons is described in hadron effective field theory by a Yukawa vertex, e.g.,\footnote{The \s\ is a parity-even scalar.  A $B B'$ final state (e.g., $\Lambda \Lambda$) has intrinsic parity +1, so its having $L=0, S=0$ is consistent with angular momentum and parity conservation and Fermi statistics.  The $\gamma_5$ in this expression results from using charge conjugation to express the matrix element in a manifestly covariant form; it does not reflect parity violation.  For details see~\cite{fw22a}.}
\be
\label{gsbb}
<\Lambda \Lambda' | \mathcal{H}_{\rm QCD}\,| S> \equiv  \frac{\tilde{g}}{\sqrt{40}}\, \bar{u}_{\Lambda'}  \gamma_5 v_{\Lambda^c} ~.
\ee
The interconversion amplitude for other baryon combinations, $\Sigma^+ \Sigma^-$, $\Sigma^- \Sigma^+$, $\Sigma^0 \Sigma^0$, $p \Xi^-$, $\Xi^- p $, $n \Xi^0$, and $\Xi^0 n $, are the same except for the sign.  The Clebsch-Gordan coefficient $\sqrt{1/40}$, derives from the requirement that the S wavefunction is fully anti-symmetrized under exchange of any two quarks.  The full wavefunction has an amplitude $\sqrt{4/5}$ to consist of products of color-octet baryons, and only $\sqrt{1/5}$ to be products of color-singlet baryons.  Within the color-singlet terms, the amplitude to have the same internal spin, flavor and color structure as two Lambda's is just $\sqrt{1/8}$, whence the $\sqrt{1/40}$.

The dynamical fusion/dissociation amplitude \gtilde\ is itself suppressed by two effects: \\ {\it (i)} The hard-core repulsion in the baryon-baryon potential reduces the wave-function overlap between the quarks in the two baryons and the quarks in the \s, which are concentrated at the origin~\cite{fzNuc03}.  \\
{\it (ii)} Interconversion between \s\ and two baryons requires the color-flavor-spin state of the quarks to be reconfigured, being three color-entangled diquarks ($ud$, $ds$, $su$) in the \s\ while being a $ud$ diquark plus a strange quark in each of the two $\Lambda$'s.  Since di-quarks are deeply bound, breaking and reconfiguring them entails a tunneling suppression.  

Figure~\ref{fig:gtilde}, reproduced from~\cite{fw22a}, gives the current experimental and observational limits on \gtilde, along with theoretical estimates for \gtilde.  The grey bands show theoretical calculations of \gtilde\  for two different assumptions for the sexaquark radius $r_S$ (entering the wavefunction overlap) and two different estimates of the tunneling suppression.  In all cases shown, the hard-core radius in the baryon-baryon potential is taken to be the standard value of 0.4 fm, but this quantity is poorly constrained and even a slightly larger value greatly reduces \gtilde~\cite{fzNuc03}.  In the upper band, $r_S$ is calculated with Eq.~\eqref{rN} and $b=0.45$, as fits baryons, while the lower band is a lower bound on the overlap taking $r_S$ to be the \s's Compton wavelength.\footnote{The wave function overlap -- Eq. (\ref{gsbb}) with $\mathcal{H}_{\rm QCD}$ replaced by unity -- was first calculated in~\cite{fzNuc03}, for free baryons and for baryons in a nucleus, using the Isgur-Karl parameterization of the spatial distribution of quarks.  That overlap calculation yielded \gtilde\ $\lesssim 10^{-6.5}$ using $r_S$ from Eq.~\eqref{rN} with $b=0.45$ and a nominal value of the hard-core radius, 0.4 fm.  The overlap decreases rapidly with increasing hard-core radius~\cite{fzNuc03} which is poorly determined; it may be $0.5$ fm~\cite{Morita+LamLamPot15} or larger, in which case the overlap $\lesssim 10^{-7.5}$.   Determining the structure of the baryon-baryon potential at short distance is not only experimentally but also theoretically difficult, because the appropriate degrees of freedom transition from baryons at long distance to quarks at short distance.  The overlap increases with increasing $r_S$, but less strongly; for further investigation of parameter sensitivities see ~\cite{fzNuc03,fw22a}. }  The overlap itself is only an upper limit on \gtilde\ since  the tunneling suppression of the transition can be very significant, as exemplified by the Gamow factor in nucleosynthesis.   The tunneling suppression is $e^{-{\rm action}}$, where the action for tunneling between configurations is estimated in Fig.~\ref{fig:gtilde} as six times the action for an individual quark to tunnel, $6(\Delta E \,\Delta t)$.  Calculating the action for each quark using $\Delta t \approx r_p/c$ and $\Delta E = $100 or 300 MeV, results in a tunneling suppression factor from 0.05 to $10^{-4}$; this range is represented by the grey bands.  The upper grey band represents the present best estimate for \gtilde.   See~\cite{fw22a} for further details.

\subsection{Experimental constraints on \gtilde\ } 
\label{sec:labgtilde}

We now briefly survey the experimental and observational constraints on \gtilde\ reported in~\cite{fw22a}.   These constraints fall into two categories:  those which apply in general, and those which only apply if sexaquarks are a significant component of dark matter.  The general constraints are shown as solid boundaries in Fig.~\ref{fig:gtilde} while dashed lines indicate those which follow if sexaquarks constitute 100\% of the DM.  Some limits are relatively independent of $m_S$, while limits based on the stability of nuclei or that of the sexaquark apply only in specific ranges of $m_S$.   

If $m_S$ is light enough, $m_S < m_D = 1876.122$ MeV, the \s\ is absolutely stable but deuterium is unstable to electron capture -- forming an \s\ with neutrino emission.   If $m_S  < m_d - m_e = 1875.101$ MeV, the deuteron nucleus can decay directly by $d\rightarrow S e^+ \nu$, although as a 3-body decay this is strongly phase-space suppressed near threshold.  The turquoise excluded region shown in Fig.~\ref{fig:gtilde} is derived~\cite{fw22a}  from the excellent limit on the rate of injection of positrons of energy greater than 5.5 MeV from the Sudbury Neutrino Observatory (SNO) detector~\cite{SNO}, placing a limit on the lifetime for $d \rightarrow S e^+ \nu_e$.   For $m_S \leq 1870$ MeV the limit is $\tau_d > 10^{28}$ yr, becoming very much stronger as $m_S$ decreases and the three-body phase space opens up;  $\tau_d > 10^{31}$ yr for $m_S \lesssim 1850$ MeV.   The turquoise excluded region in Fig.~\ref{fig:gtilde} gives the corresponding constraint on \gtilde.   Deuteron stability disfavors $m_S \lesssim 1850$ MeV but does not totally exclude it, in the absence of stronger theoretical  confidence in calculating \gtilde.  The orange dashed line is the earlier, weaker limit from~\cite{fzNuc03} based on the SuperK background level compatible with $^{16}$O instability.\footnote{Reference~\cite{gross+strumia+18}, using the results of~\cite{fzNuc03} but employing a wavefunction that does not properly reflect the known short-distance suppression of the baryon-baryon wavefunction in nuclei, exaggerates the constraining power of $^{16}$O stability.}

If $m_S>1870$ MeV, the strongest secure constraint on \gtilde\ which does not rely on sexaquarks comprising the DM, follows from the requirement that the formation time of the \s\ in a doubly-strange hypernucleus be longer than the lifetime of the $\Lambda$~\cite{fzNuc03}.  This bound is shown as the green excluded region in Fig.~\ref{fig:gtilde} and is evidently too weak to exclude values of \gtilde\ expected for a compact sexaquark.\footnote{Placing a limit on \gtilde\ from the formation time constraint if $m_S > 2 m_\Lambda - m_\pi = 2090$ MeV requires more detailed treatment of nuclear effects than provided in Ref.~\cite{fzNuc03}, hence the phasing-out of the constraint above this mass.  With further work modeling the nuclear physics, the hypernuclear constraint could be extended to higher mass.}   A stronger limit is possible if the arrival time distribution of  neutrinos from SN1987a can be used to constrain the cooling time of proto-neutron stars (as suggested in~\cite{raffeltSN1987a} but questioned in~\cite{blum+SN1987a_20}).   Under the assumption that use of SN1987a is valid, the authors of Ref.~\cite{mcdermott+SNe19} calculate the effect of an additional process $\Lambda \Lambda \rightarrow S + \gamma$ on the cooling rate of SN1987a.  They find that requiring \s\ production to not equilibrate faster than 10 s implies that $< \sigma_{\Lambda \Lambda \rightarrow S \gamma} \, v>\, \lesssim 4 \times 10^{-34}\,{\rm cm^3/s}$.  Equation (8) of Ref.~\cite{mcdermott+SNe19} expresses $< \sigma_{\Lambda \Lambda \rightarrow S \gamma} \, v>$ in terms of \gtilde\ (=$\sqrt{40} g_{S \Lambda}$ in their notation), the binding energy $B_S = 2 m_\Lambda - m_S$ and the temperature, taken to be 30 MeV.  Solving for \gtilde\ gives the limit shown in the red dotted line in Fig.~\ref{fig:gtilde}.\footnote{Reference~\cite{mcdermott+SNe19} does not actually report their constraint on $g_{S \Lambda}$, but instead reports limits on $r_S$ through an overlap calculation.  As the $\Lambda \Lambda$ wavefunctions used are not given, we cannot compare to that aspect of their work.}  

The sexaquark mass range $1870 ~{\rm MeV}< m_S <  2 m_n = 1879$ MeV is special.  In this range the constraints on \gtilde\ are much weaker because \s\  is massive enough that deuteron decay in SNO would not produce a detectable positron signal, yet the \s\ is essentially absolutely stable because the phase space for $S \rightarrow d e^- \bar{\nu}$ is so small.   In this region, the only constraints are those from hypernuclear decay and SN1987a.

\subsection{Dark Matter constraints on \gtilde\ } 
\label{sec:subgtilde}

If the \s\ 
is a major component of the dark matter, several stronger constraints on \gtilde\ in the higher mass range can be derived~\cite{fw22a}.  The condition that the lifetime of the \s\ be longer than the age of the Universe,  $\tau_S >  \tau_{\rm Univ}$, is shown in Fig.~\ref{fig:gtilde} in blue; it is readily satisfied for $m_S < 2054$ MeV because below this mass $S \rightarrow \Lambda n$ is  kinematically forbidden and the \s\ decay amplitude is doubly-weak~\cite{fzNuc03}.   We can calculate the amount of deuterium produced by DM decays since the epoch of primordial nucleosynthesis as a function of \gtilde, and compare that to the difference between the predicted deuterium abundance from Big Bang nucleosynthesis and the value measured at late time in Ly-$\alpha$ systems ($z \sim 4$).  Assuming the \s\ accounts for all of the dark matter, this excludes \gtilde\ in the  brown region in the upper right corner~\cite{fw22a}.  Essentially the same region is excluded by requiring the DM to baryon ratio implied by primordial nucleosynthesis when the Universe was a few minutes old, to agree with the value inferred from the CMB when the Universe was 300,000 years old~\cite{fwx_SDM20}.  

The strongest constraint on \gtilde\ for an unstable \s\ comprising all of the dark matter comes from limits on heating of astrophysical systems.  For $m_S > m_n + m_\Lambda$ the process is $S\rightarrow \Lambda n$ with the $\Lambda$ decaying predominantly to nucleon and pion, while for $2 m_n < m_S < m_n + m_\Lambda$ the process is the doubly-weak $S\rightarrow nn$.  In both cases, the final step is $n \rightarrow p e^- \nu$.  Following~\cite{ww21,wf21}, this excludes the magenta region~\cite{fw22a}.  The heating constraint can be simply rescaled from Fig.~\ref{fig:gtilde} if the fraction of DM composed of sexaquarks is not 100\%.  For instance if \s's only account for 1\% of the DM, the rate of  $S\rightarrow nn$ could be 100 times higher and the upper bound on \gtilde\ would increase by a factor 10.

The horizontal black dashed line in Fig.~\ref{fig:gtilde} is the maximum value of \gtilde\ consistent with survival of \sdm\ as the Universe transitions from the quark gluon plasma into a hot hadronic plasma ~\cite{fudsDM18}.  This value is calculated by requiring that the reaction rate of breakup processes such as $K^+ S \rightarrow p\, \Lambda$  is less than the expansion rate of the Universe at $T \approx 150$ MeV.  The survival condition becomes easier and easier to satisfy as the Universe cools, and the upper bound on \gtilde\ would be weakened if there were continuing production of \sdm\ throughout the QGP-hadron transition.  Considering all of the relevant reactions, survival of \sdm\ is guaranteed if \gtilde\ is less than $\lesssim 2 \times 10^{-6}$~\cite{fDMtoB18}.  This constraint is satisfied based on overlap alone, even without taking into account tunneling suppression.   Hence the conclusion of \cite{kt18} that dark matter cannot be di-baryonic, while applicable for a loosely-bound H-dibaryon, is inapplicable to \sdm.  

To summarize this section, the QCD interconversion amplitude \gtilde\ between \s\ and two baryons must be small, as shown in Fig.~\ref{fig:gtilde}.  Theoretical estimates based on wavefunction overlap are very sensitive to the assumed sexaquark radius, with a value of $\tilde{g} \lesssim 10^{-7}$ being obtained as a fiducial value.  The observational constraints on \gtilde\ are strongest below $m_S \approx 1870$ MeV;  for higher masses $\tilde{g} \approx 10^{-4.5}$ is compatible with experimental constraints as long as dark matter does not predominantly consist of sexaquarks.  A small \gtilde\ makes calculating the sexaquark mass in lattice QCD very challenging.  The HAL QCD strategy is to reconstruct the $\Lambda \Lambda$ potential, then calculate bound state masses in that potential.  However a small \gtilde\ implies that accurately probing the $\Lambda \Lambda$ potential at small distances through $\Lambda \Lambda$ scattering will be difficult.  

\section{Astrophysical and related constraints} 
\label{sec:astro}

If a sexaquark exists with $m_S \lesssim 2.05$ GeV, sexaquarks can naturally be the dark matter~\cite{fDMtoB18}. In this section we briefly report constraints on sexaquark dark matter (SDM) and some implications.

If the dark matter is composed all or in part of sexaquarks, the possibility of a non-negligible scattering cross section with baryons can have observable astrophysical and cosmological consequences.  References~\cite{xf21a,xf21b} provide a comprehensive compendium of present limits on DM-baryon interactions for DM in the GeV mass range.  A full non-perturbative analysis  is essential, solving the Schr\"odinger equation for extended nuclear sources rather than assuming Born approximation scaling with nuclear mass number $A$.  Cosmological limits from small scale structure in the CMB give a limit on the dark matter nucleon cross section, $\sigma_{Xp} \lesssim 10^{-24.5}\, {\rm cm}^2$, and a similar limit follows from cooling of gas clouds~\cite{wf21,xf21a}.  Recently, these limits have been improved by combining constraints from big bang nucleosynthesis with bounds from a novel dewar experiment~\cite{nfm18,nbn19}, giving $\sigma_{Xp} \lesssim 10^{-26.2} \,{\rm cm^2}$~\cite{xf21b} for $m_S \approx 2$ GeV.  This is at the upper range of plausible \s-nucleon cross sections (Sec.~\ref{sec:prop-xcn}).\footnote{Direct DM detection experiments are at much lower energy than encountered in accelerator experiment contexts, with Galactic DM having a typical 300 km/s velocity implying keV-range kinetic energies for \sdm.  Furthermore, direct detection experiments with significant material overburden lose sensitivity when the DM particles reaching the detector have lost enough energy via scattering in the overburden that they are no longer capable of triggering the detector.   This effect is particularly important for DM in the GeV range; as a consequence, the ultra-sensitive deep underground WIMP detectors are not sensitive to sexaquark DM.  Near-surface experiments to date suffer from the problem that \sdm\ deposits little energy in detectors and the detector responses have not been calibrated in the relevant regime~\cite{mfVelDep18,xf21a}.  Note that SENSEI~\cite{SENSEI18}, which is sensitive to GeV-and-below masses, relies on a DM-electron coupling which is not present for the electrically neutral \s.}  
 
A DM-baryon interaction in the $10^{-26} {\rm cm^2}$ range would have implications for astrophysics.  A recent detailed analysis~\cite{lf_sparc21} of the rotation curves of 121 well-measured galaxies in the SPARC dataset found that inclusion of a (likely puffy) DM disk in addition to a spherical halo improved the fit, possibly pointing to a DM-baryon interaction as present for \sdm.   A DM-baryon interaction could also help solve the core-cusp problem of standard cold dark matter halos, although baryonic physics alone may be sufficient.\footnote{The core-cusp problem motivated the suggestion of Self-Interacting Dark Matter~\cite{ss:SIDM}. \sdm\ has self-interactions due to $\phi_0$ exchange, but the analysis of~\cite{fwx_SDM20} predicting $\sigma_{SS}$ in terms of the \s-nucleon Yukawa coupling and the nucleon-$\phi_0$ coupling taken from hadron effective field theory fits, combined with the new limit $\alpha_{SN} \lesssim 0.8$ obtained in Ref.~\cite{xf21b}, implies $\sigma_{SS} \lesssim 0.02 \, {\rm cm^2 /g}$ -- much smaller than invoked in SIDM models and easily compatible with limits from the Bullet Cluster, e.g.,~\cite{TulinYuPhysRpts17,RMEbulletSIDMlim17}.} 
As a final example, we note that the cooling-flow catastrophe in X-ray
clusters can be eliminated by mild heating of the intra-cluster gas induced by DM-gas interactions~\cite{QinWu01,chuzNusser06,wfHIDM18}.   

A potential astrophysical challenge to the existence of a stable
sexaquark is the observation of neutron stars with masses
above 2 $M_\odot$. Hyperons, which we know exist, present a similar problem 
for the existence of massive neutron stars, a problem resolved by quark deconfinement ~\cite{baym+NSrev18,kojoBaymHatsudaNS21} at densities below the emergence of a hyperon-dominated phase.   
Reference~\cite{shahrbaf+22} takes an empirical approach to examining the compatibility of neutron star observations with the existence of a sexaquark, allowing for both a deconfined phase and a density dependence of the hadron masses as in~\cite{marques+DD2YT17}, and demanding  simultaneous consistency with both the GW170817 constraints on tidal deformability~\cite{TidalDefGW_18} and with the mass-radius relation of pulsars including the highly constraining NICER results on PSR J0740+6620~\cite{NICER_0740_21,rileyNICER21,millerNICER21}.  Reference~\cite{shahrbaf+22} finds that:  (1) a deconfined (quark matter) core is essential to support the most massive neutron stars; (2) a significant contribution of hyperons must be excluded in any density regime, as that excessively softens the equation of state; (3) present observational constraints are compatible either with early deconfinement such that sexaquarks also never play a role, or with a sexaquark-dominated phase.   This raises the interesting possibility of low mass neutron stars with a sexaquark core, or a more massive neutron star with a sexaquark layer, potentially giving a means to probe and constrain sexaquark properties~\cite{shahrbaf+22}.   

\section{Strategies for accelerator discovery}
\label{sec:expts}

The cross section for production of \s\ in exclusive hadronic collisions is proportional to $\tilde{g}^2$, so that \s\ production in reactions near threshold such as $K^- p \rightarrow S \bar{\Lambda}$ is severely suppressed compared to competing processes without \s\ production (contrary to expectations for the more weakly bound H-dibaryon or a di-$\Lambda$ molecule).   Thus search strategies which are not suppressed by $\tilde{g}^2$ are needed.   Furthermore, a good search strategy will have some feature of the observation or measurement that provides a clear demonstration of -- or smoking gun for -- production of a new particle.   The general experimental challenge to discovering an \s\ and demonstrating its existence is that \s's are neutral and similar in mass to neutrons, but interact less and are much less abundant.   


The sexaquark production rate in hadronic interactions can be expected to vary strongly with the experimental conditions.  At low energies where production is through exclusive processes, the production fraction is reduced relative to production of other hadrons by a factor $\tilde{g}^2$, which at low momentum is measured to be $ \lesssim 10^{-9}$ and estimated to be $ \lesssim 10^{-12}$ (see Sec.~\ref{sec:gtilde} and Fig.~\ref{fig:gtilde}).  As the relative momentum of the \s\ and baryons increases, the simple overlap calculation of~\cite{fzNuc03} is no longer applicable and the effective \gtilde\ may increase.  At still higher relative momentum, when \s\ production occurs directly from a multi-quark state such as found in the central region of a high energy collision or in $\Upsilon(1S,2S,3S)$ decay, \gtilde\ does not enter and the rate can be higher.  In high-multiplicity inclusive reactions such as $p p$ collisions at the LHC,  the production rate can be estimated by extending the heuristic that there is a price of $\mathcal{O}(10^{-1})$ for each additional quark in the state, based on the meson to baryon ratio in the central region of high energy collisions and $Z$ decay.  If applicable, this heuristic suggests the production rate of \s\ (and \sbar) in the central region of $pp$ interactions is of order $ 10^{-3}$ to $10^{-5}$ relative to neutrons, depending on whether a factor-10 penalty applies just for the quarks in the \s\ or also for those in the accompanying anti-baryons.  

Even with a production rate relative to neutrons of order  $10^{-5}$, \s's would be routinely produced in the LHC since on average about 0.1 neutrons are produced in the central region of a $pp$ collision, per unit of rapidity.\footnote{See \cite{CMSchgpartmult10} for charged particle multiplicities in the central region of pp collisions at $\sqrt{s}= 0.9,\, 2.76$ and 7 TeV.  The $dN/dY$ values are slowly varying with c.m. energy, and $p$ and $\bar{p}$ are nearly equal, so to adequate approximation we take them equal, and equal to $n$ and $\bar{n}$, based on isospin invariance. For reference, the ATLAS inner tracker covers a pseudorapidity range $\pm 2.5$.} The inelastic $pp$ cross section at 13 TeV is about 80 mb~\cite{inelppATLAS16} and the integrated $pp$ luminosity at 13 TeV to date is $\approx 150 \, {\rm fb}^{-1}$.  Thus about $10^{16}$ inelastic $pp$ collisions have occurred, which could have produced $\gtrsim 10^{10}$ \s\ and \sbar's  in the central rapidity region.  

Heavy ion collisions are also very attractive for searching for \s\ and \sbar, in part because the production rate in central collisions is large, but even more importantly because the \s\ and \sbar\ production rate can be estimated more confidently than for $pp$ collisions, as discussed in Sec.~\ref{ss:HIC} below.  The cross section for the 50\% most central Pb-Pb collisions at 5 TeV CM is about 4 b~\cite{ALICErefxcn13} and the integrated luminosity collected to date is a few $ {\rm nb}^{-1}$, implying $\gtrsim 10^{10}$ central Pb-Pb collisions.   With upcoming high luminosity running, these numbers will increase by large factors.   

While the number of \s\ and \sbar's produced may be large, their presence would not have been obvious in experiments to date.  Unlike in a search for heavy Beyond the Standard Model neutral particles, e.g., as expected in supersymmetry, where a new particle would escape with large missing energy or missing transverse momentum~\cite{f:23}, the \s\ has a typical QCD transverse momentum, $\mathcal{O}(1)$ GeV, so the energy or momentum imbalance in an LHC detector -- where the particle multiplicities are very large and many particles are not detected -- would not be significant.   Moreover, neutrons with an abundance $\mathcal{O}(10^{2.5}\!-\!10^5)$ greater than \s's, constitute a large background with very similar properties.  Both neutron and \s\ have zero charge, inhabit the same mass range, and have relatively low interaction probability in most detectors.  The rare interaction of an \s\ or \sbar\ would normally be dismissed as the occasional interaction of a neutron or anti-neutron, if indeed it was even noticed.   

In the following we discuss several search strategies, each with their own strengths and limitations.  The approaches are complementary in that one may be quickest to establish the existence of a new neutral stable particle with properties consistent with the \s\ and measure the product of its production and scattering cross section, while another could demonstrate the characteristic baryon number and strangeness quantum numbers.  Followup experiments could determine the mass and the \s-baryon interconversion parameter, \gtilde.   

\subsection{Previous searches and low energy exclusive reactions}
\label{ss:LEX}

In Ref.~\cite{fS17} we comprehensively reviewed experimental searches for the H-dibaryon and determined that searches up to then (2017) either were only sensitive to masses above 2 GeV, or searched for decay products and hence were not capable of discovering a stable \s, or were not sensitive enough to discover the \s.  
A closer examination of the literature shows that sexaquarks would have escaped detection over the entire mass range in which they are effectively stable ($m_S < 2054$ MeV).  Appendix~\ref{app:otherexpts} gives more details.

A key approach of the H-dibaryon searches was to use a low energy exclusive reaction such as $K^- p \rightarrow  S \bar{\Lambda}$, in which the baryon number and strangeness of the unseen \s\ (H) would be clear, and the missing mass would give $m_S$.  Just a few unambiguous events like this would constitute a discovery if the resolution is good and all particles are identified.  But in exclusive or low-multiplicity interactions at relatively low energy, the production cross section is $ \sim \! \tilde{g}^2$ times the cross section for a kinematically similar process not involving \s.   So given the upper limits on \gtilde, the rate for $K^- p \rightarrow  S \bar{\Lambda}$ would be at best $ 10^{8}$ times lower than for other channels not including the \s.  The required degree of background rejection is currently  not achievable.

\subsection{Final states of $\Upsilon(1S,2S,3S)$ decays}
\label{ss:ups}

The states $\Upsilon(1S,2S,3S)$ decay to hadrons through an intermediate three-gluon state.  The gluons then convert to quarks and anti-quarks, which subsequently form hadrons.   Since the characteristic size of the $ggg$ state from which the final hadrons emerge is (10 GeV)$^{-1}$= 0.02 fm, \s\ and \sbar\ are produced directly without the factor \gtilde\ which suppresses exclusive scattering reactions.\footnote{The Belle collaboration recognized early on that $\Upsilon$ decay final states are potentially enriched in flavor singlets, in particular a possible H-dibaryon~\cite{BelleHdibaryon13}.   However their search required evidence of H-dibaryon decay into $\Lambda$ final states, so would not have seen a stable sexaquark.}

Two distinct approaches using $\Upsilon(1S,2S,3S)$ decays are discussed below: {\it (i)} reconstruction of the missing mass for final states consistent with an escaping \s\ or \sbar, 
and {\it (ii)}  demonstration of a correlated, statistically significant imbalance in the baryon number and strangeness of observed final particles in the totality of $\Upsilon(1S,2S,3S)$ decays.  

An ideal reaction for the first approach, proposed in~\cite{fS17} is
\be
\label{eq:ups}
\Upsilon~~[ \rightarrow {\rm gluons}] \rightarrow S \, \bar{\Lambda} \, \bar{\Lambda}~~{\rm or} ~\bar{S} \, \Lambda \, \Lambda~~ + {\rm pions~ or} ~\gamma ~,
\ee
(or both pions and $\gamma$) with the \s\ or \sbar\ escaping undetected.   The second approach, developed below, exploits the fact that an \s\ or \sbar\ carries $ |B-S | = 4$.  Since $B$ and $S$ are conserved in the strong interactions and the $\Upsilon$ has $B=S=0$,  the observed final state when an \s\ is produced and escapes would have a net $B=-2,\, S=+2$ and $ B-S = -4$, and the opposite for an escaping \sbar.

The experimental requirements for a significant signal in either approach depends on the branching fraction $ \mathcal{F}$ for {\it inclusive} \s\ plus \sbar\ production.  For a search utilizing a particular exclusive final state, the sensitivity further depends on the exclusive/inclusive ratio for that channel.  In an inclusive short-distance-initiated reaction like decay of $\Upsilon(1S,2S,3S)$ the closure approximation is valid and estimation of $ \mathcal{F}$ for inclusive \s\ or \sbar\ production should be fairly robust.  The calculation, detailed in Appendix \ref{app:UpsStat}, yields $ \mathcal{F} \approx 2.7 \times 10^{-7}$.  The small value is due to the necessity of producing three extra gluons to have the required minimum six $q \bar{q}$ pairs, at a penalty of $\alpha_s^6$ in the rate, the low probability of six quarks or antiquarks being nearest neighbors, and the low probability of six quarks or antiquarks of the required flavors being in a color-flavor-spin singlet state.    The branching fraction for producing \s\ and \sbar\ in the continuum region between the $\Upsilon(1S,2S,3S)$'s and on the $\Upsilon(4S)$ is significantly less than on the $\Upsilon(1S,2S,3S)$ resonances because then the final state is initialized by $q \bar{q}$ rather than three gluons, requiring still higher orders of $\alpha_s$ to form the $6q 6 \bar{q}$ final state.

If all the accompanying final particles are seen, as in processes like (\ref{eq:ups}), the mass of the unseen \s\ or \sbar\ can be reconstructed from 4-momentum conservation:  $m_S^2 = (p_\Upsilon - p_{\Lambda 1}- p_{\Lambda 2} - \Sigma p_{\pi's\&\gamma's})^2$.  The resolution is so good in some detectors, $\mathcal{O}(20$ MeV), that even a handful of events appearing to be $\bar{\Lambda} \bar{\Lambda}$ or $\Lambda \Lambda $ + pions and/or gammas, consistent with having a common missing mass, would be a powerful smoking gun for the existence  of the \s\, and could determine or strongly constrain its mass.   
Other final states besides $\Lambda \Lambda$/$\bar{\Lambda} \bar{\Lambda}$ are also discovery avenues, e.g., $\Xi^- p$ instead of $\Lambda \Lambda$, or a $\Lambda $ can be replaced by $K^- p $.   As long as no baryon number or strangeness carrying particle escapes detection besides the \s\ or \sbar, any combination of hyperons and mesons with B= $\pm$ 2, S= $\mp$ 2 quantum numbers, including final states with higher multiplicities, can be used.    
The $\bar{\Lambda} \bar{\Lambda}$ and $\Lambda \Lambda $ final states are very good because the $\Lambda $'s short decay length ($c \tau = 8$ cm) and 64\% branching fraction to the 2-body charged final state $p \, \pi^- $, mean $\Lambda$'s and $ \bar{\Lambda}$'s can be reconstructed with high efficiency and their 4-momenta can be well-measured.   The drawback of this approach is that it is difficult to obtain an adequate dataset of events in which all final particles are identified and the final state is known not to have missing baryons or strange particles.   

BABAR performed a search for {\it exclusive} \s\ and \sbar\ production, accompanied only by $\bar{\Lambda} \bar{\Lambda}$ or $\Lambda \Lambda$, and placed an upper limit on the branching fraction 
BF$_{S\bar{\Lambda}\bar{\Lambda} + \rm h.c.} < (1.2-2.4) \times 10^{-7}$ \cite{babarS18}.   To compare this to expectations, Appendix \ref{app:UpsExc} estimates the penalty for demanding an exclusive final state, by examining exclusive branching fractions for other channels in $\Upsilon$ decay.  The exclusive penalty is found to be at least a factor $10^{-4}$.  Thus Babar's sensitivity in the search \cite{babarS18},  is by far insufficient to shed light on the possible existence of a stable \s.  A more hermetic detector such as Belle-II could  carry out a semi-inclusive search as indicated in (\ref{eq:ups}), which would be sensitive to a much higher portion of \s\ and \sbar's.  

A complementary strategy for demonstrating production of \s\ and \sbar's in Upsilon decay is to measure the proportion of events having specified numbers $\{N_{\rm B}, N_{\rm S}, N_{\bar{\rm B}}, N_{\bar{\rm S}}\}$ of baryons, strangeness +1 particles, anti-baryons and strangeness -1 particles in the final state, to establish a statistically significant excess of events with the correlated B-S = $\pm 4$ accompanying \s\ and \sbar\ production.  The  feasibility of this approach depends on $\mathcal{F}$ (the inclusive branching fraction of \s\ or \sbar), $N_{\rm tot}$,  the total number of $\Upsilon(1S,2S,3S)$ decays recorded, and the identification efficiency of the various baryons and strange particles, including losses from less than $4 \pi$ detector coverage.  

\begin{figure}
 \centering
 \includegraphics[width=0.68 \textwidth]{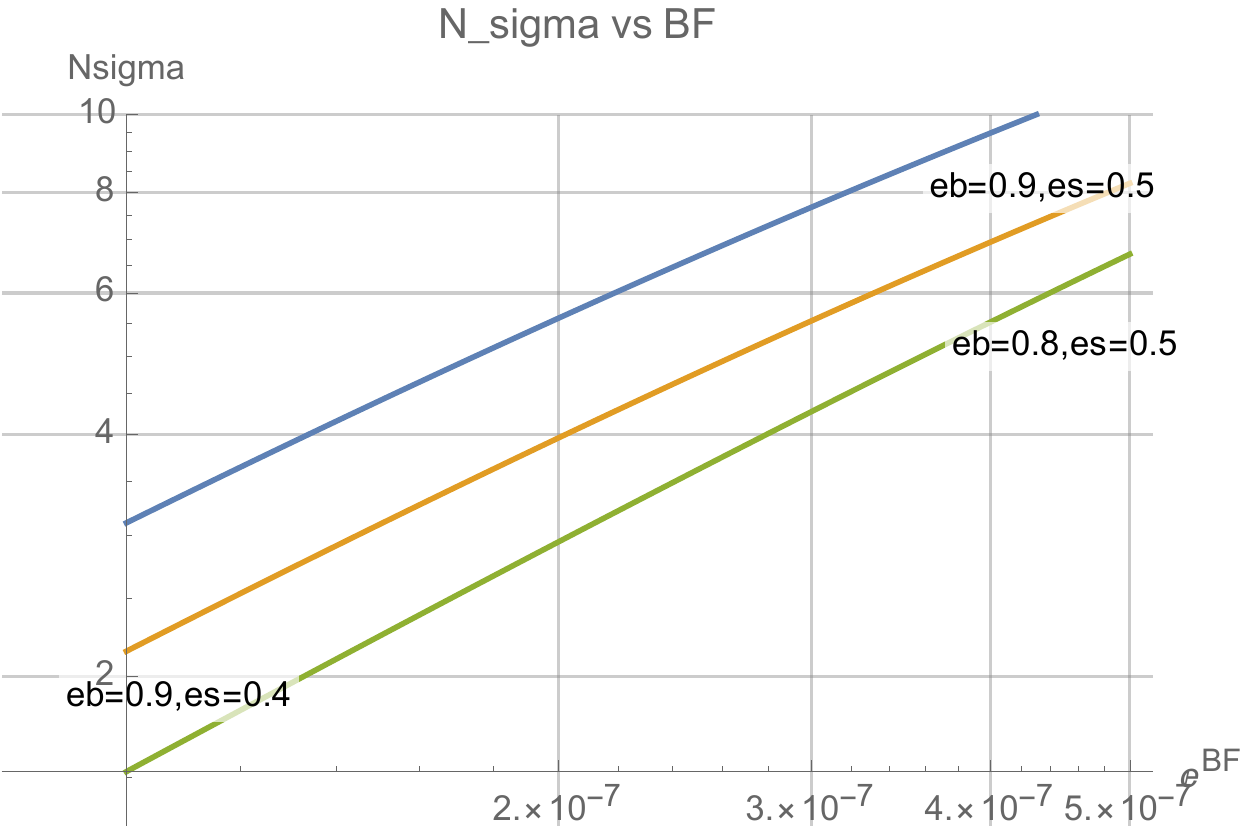}
 \caption{Estimated significance of a $|B-S| \neq 0$ signal for $10^9$ $\Upsilon(1S,2S,3S)$ decays as a function of sexaquark branching fraction, $ \mathcal{F}$, for several representative baryon and strangeness ID efficiencies, from top to bottom $\{e_b,e_s\} = \{0.9,0.5\}, \{0.9,0.4\}, \{0.8,0.5\}$.} \label{fig:NsigvsBF}
\end{figure}

An approximate method to assess the possible sensitivity of this approach is described in Appendix~\ref{app:B-S=4}, using a single effective efficiency for identifying baryons and anti-baryons, $e_{b}$, and similarly $e_{s}$ for strange and anti-strange particles.  The Belle-II particle ID efficiency for baryon and kaons is 0.8-0.9 depending on how hard they cut, which depends on the backgrounds they want to suppress, so we show results for $e_b = 0.8$ and $0.9$.  Hyperons have clearly determined strangeness, but only half the kaons do (since the neutral $K_0^{L,S}$ mass eigenstates are superpositions of $K_0$ and $\bar{K_0}$) so we report results for $e_s=0.4$ and $0.5$.  Figure~\ref{fig:NsigvsBF} shows the statistical significance by which a $|\rm{B-S}|=4$ excess can be established as a function of the sexaquark inclusive branching fraction  $\mathcal{F}$, for $N_{\rm tot} =10^9$ $\Upsilon$ decays and three different combinations of assumed efficiencies $e_b$ and $ e_s$;  the significance scales as $\sqrt{N_{\rm tot}}$.   Not surprisingly, large $e_b$ is more important than large $e_s$, since baryons are rarer and thus any sexaquark contribution makes a larger relative impact on the baryon abundances.   A sensitivity of 4-8$\,\sigma$ appears to be possible with Belle-II, depending on the actual effective efficiencies, for the estimated inclusive branching fraction $ \mathcal{F} \approx 2.7 \times 10^{-7}$ and $10^9 ~ \Upsilon(1S,2S,3S)$ decays.  

The sensitivity to \s\ production can be increased by exploiting events with final states with other values of detected $|\rm{B-S}|\neq 4$.  For instance, events with two baryons and one negative strangeness particle  have some sensitivity to $\mathcal{F}$ but with a larger fractional contribution from standard channels, $ \{2b\,1\bar{s}\}$ in the notation of  Appendix~\ref{app:B-S=4}. Cases like $ \{ 1b\,1\bar{b}\,1s\,1\bar{s}\}$  which get no contribution from \s\ or \sbar\ are valuable to develop confidence that the systematics of the background are fully understood.
A more accurate forecast of the sensitivity of this strategy at Belle-II requires a real detector simulation.   This is facilitated by the hadronic event generator EPOS-LHC having been modified to incorporate \s\ and \sbar\ production in $\Upsilon(1S,2S,3S)$ decay as well as $pp$ and heavy ion collisions~\cite{pierog+22}. 

\subsection{High intensity photon beams}
\label{ss:Jlab}
J-Lab has a tagged photon beam of energy 9 to 12 GeV, with $10^8$ photons/second on target.  The GlueX experiment, which anticipates collecting of order $10^{12}$ interactions, has adequate kinematic reach to probe reactions such as
\be
\label{eq:gluex}
 \gamma \, p \rightarrow S \, \bar{\Lambda} \, K^+ + pions ~.
 \ee  
A 12 GeV photon provides $E_{\rm CM} = 4.84$ GeV, which is 1.35 GeV above the 3.5 GeV total mass of $S, \, \bar{\Lambda}, \, K^+$ for  $m_S = 2 m_p$, leaving room to spare for phase space and pion production.   Given the potentially very large number of events, and depending on the solid angle coverage and tagging efficiency, good discovery channels could be either the missing mass structure or unbalanced baryon number and strangeness due to an escaping \s. The cross-section for the process~\eqref{eq:gluex} is suppressed relative to other final states by a factor $\tilde{g}_{\rm eff}^2$ which should not be so small as $\tilde{g}^2$ discussed in Sec.~\ref{sec:gtilde} since the momentum transfer to the \s\ is larger here.

\subsection{Heavy Ion Collisions}
\label{ss:HIC}
Central relativistic heavy ion collisions are a very attractive production site for \s\ and \sbar, despite the integrated luminosity for heavy ions at the LHC being much less than for $pp$, for two reasons.  Firstly, production of heavy or complex states is less strongly suppressed than in other processes, as discussed below.  Secondly, \s\ and \sbar\ production in central relativistic heavy ion collisions should have some relationship to production in the Early Universe, if sexaquarks are indeed the DM.  The processes are not identical because in the Early Universe the cooling timescale at the hadronization transition is $\sim 10^{-5}$ s and the medium is infinite, while in a heavy ion collision the cooling time is very much shorter and the plasma expands into the vacuum.  Nonetheless, measurement of (or limits on) \s\ and \sbar\ production in a heavy ion collision will be very informative.

With regard to production,  \cite{Andronic+HIC_Nature18} obtains an excellent fit to the relative abundances of final particles observed by ALICE in central Pb-Pb collisions (including such complex and exotic states as anti-tritium), under the assumption of statistical equilibrium at a temperature $T=156$ MeV and accounting for production and decays of resonances.   The main systematic uncertainty in their fit is associated with treatment of the resonances.  A similar approach applied to \s\ and \sbar\ production should give a result similar to deuteron and anti-deuteron production:  $dN/dY \approx 10^{-1}$ in the central Pb-Pb collisions at $\sqrt{s_{NN}} = 2.76$ TeV -- about a factor-300 lower than $p$ and $ \bar{p}$ production.  A more detailed coalescence model for sexaquark production in relativistic heavy ion collisions predicts a production rate relative to deuterons of 1.4 to 0.22 for $m_S$ from 1700 to 1950 MeV \cite{blaschkeKabana+HIC21}.

The strategy outlined in Sec.~\ref{ss:ups} in the context of $\Upsilon(1S,2S,3S)$ decays --- looking for an excess of events in which the observed final state particles have $ | \, B - S | = 4$ due to production and escape of an \s\ or \sbar\ whose baryon number and strangeness is balanced by the observed final state hadrons --- is not an effective approach for heavy ion or LHC p-p collisions.   One problem is the impossibility of perfectly measuring the B and S of each final particle given the very high CM energy and multiplicities.  A further problem is the limited rapidity range that can be observed.  For example in Ref. \cite{ALICE_Bflucs_19}, ALICE presents a study of the event-to-event fluctuations in the baryon number of particles with $0.6 < p < 1.5$ GeV/c and $| \eta | < 0.8$.  They find that for central collisions the difference in number of baryons and anti-baryons is of order the sum.  With such large fluctuations, it would be difficult to discern a population of events above background with $B-S = \pm 4$, unless a large portion of the final particles can be identified.  A detector simulation would be needed to properly assess the prospects.

However the long-interaction-length neutral particle technique discussed in the next section could be very effectively used for  central relativistic heavy ion collisions.

\subsection{Search for long-interaction-length stable neutral particle}
\label{ss:LIL}
The challenge in searching for inclusive \s\ and \sbar\ production in a high energy collision is identifying them in the face of vastly more neutrons, as mentioned earlier.   The task is actually more difficult than discovery of new massive beyond-the-standard-model particles such as massive supersymmetric particles, where decay to a neutral stable particle  (lightest supersymmetric particle) which escapes undetected provides a missing-energy or missing-momentum signature~\cite{f:23}.   Charged massive long-lived particles can be detected by their characteristic energy deposition, and time-of-flight is also useful.  Recent interest in milli-charged particles and dark photons kinematically mixed with ordinary photons, has stimulated efforts such as MilliQan~\cite{milliQan21} looking for a component of penetrating, weakly-ionizing particles.  Experiments such as FASER~\cite{FASER21} search for light, weakly interacting particles produced in the forward direction which can penetrate 100's of meters of shielding and decay in the detector.   Here, we outline requirements for searching inclusively for a neutral, stable but moderately interacting particle in the GeV mass range, where the major challenge is discriminating it from neutrons.

\begin{figure}[t]
	\centering
		\includegraphics[trim = 0.3in 4.2in 3.0in 2.1in, clip, width=0.93\textwidth]{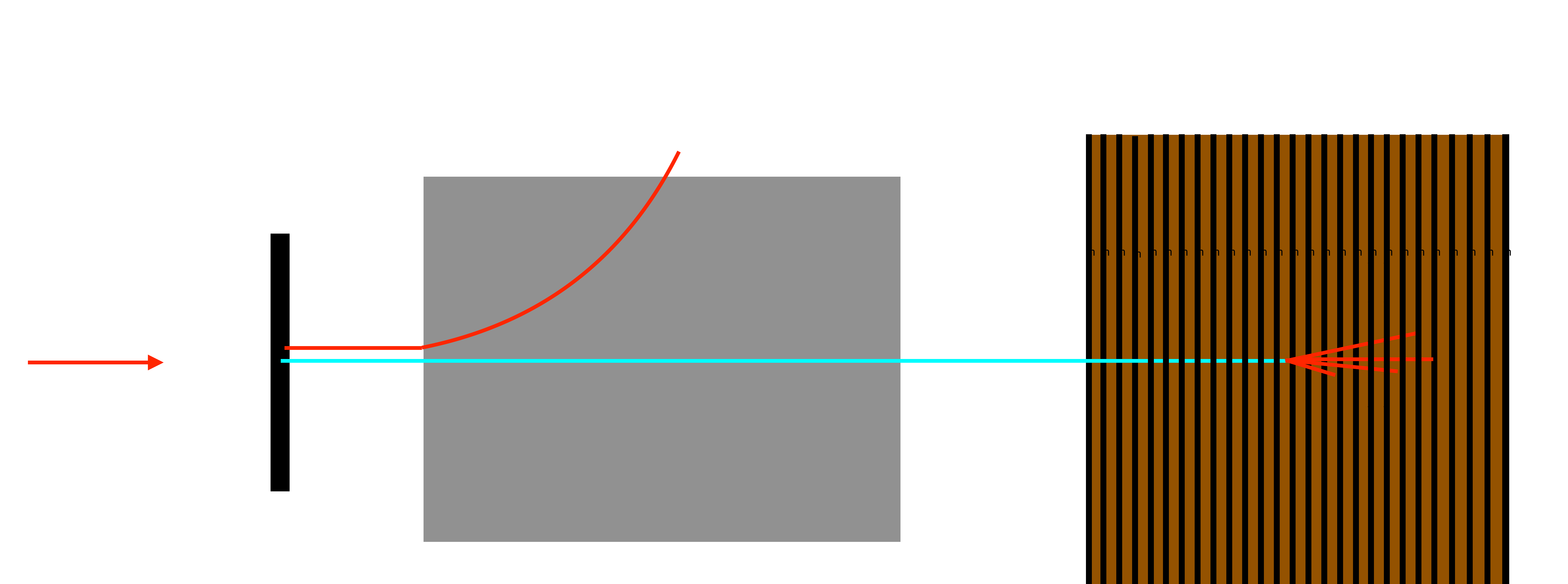}
	\caption{Schematic (not to scale) of possible realizations of a long-interaction-length neutral particle detector.  The key element is the brown block which represents Fe or other absorber, with interleaved detectors (black) to localize vertices and to identify neutral particle interaction points and decay vertices of residual $K^0_L$'s.   
The cyan lines represent neutrons or S's (and their antiparticles) and the red lines indicate charged particles.  In a fixed-target configuration, the thin black rectangle on the left is the target and the grey block is absorber and/or sweeping magnet.  In a collider configuration, the red incoming arrow should be ignored and the black rectangle represents the collision region.  The grey block could be the inner detector of an existing experiment; in a dedicated apparatus its function is to remove everything but n's and S's as efficiently as possible. \label{fig:LILN}}
  \end{figure}

A possible strategy is to search for evidence of a non-decaying neutral component  different from known neutral long-lived particles, with an interaction length longer than that of neutrons.  Figure~\ref{fig:LILN} illustrates schematically the general strategy to search for an anomalous component of long-interaction-length neutral  stable particles.  Due to the small value of $\tilde{g}$, the \sbar\ annihilation channel is much smaller than its scattering channel, so \sbar\ interactions can be taken to be indistinguishable from \s\ interactions in this context. 
In one realization of an experimental setup, a target is followed by a sweeping magnet and decay region to remove charged and short-lived neutral secondaries of the high energy collision. (Additional passive absorber, not shown, might be employed to reduce $K_L^0$'s and other known particles.)  This region is followed by an instrumented region with particle tracking interleaved with absorber, indicated in black and brown, respectively.  The purpose of the tracking layers is to measure the longitudinal position, $z$, of $n$, $\bar{n}$, \s\ and \sbar-initiated scattering or annihilation events, and to reject decays or interactions of long-lived neutrals.   An alternate realization is to use an existing LHC detector to identify the charged particles and neutrals that decay within it, with the layered tracker/absorber being outside.   Or, conceivably, an existing detector component may be suitable for this purpose. Note that in the simplest setup, the long-interaction-length neutral (LILN) detector only records the position of the interaction and rejects background; no discrimination between an $n$, $\bar{n}$, \s\ or \sbar\ induced interaction is required.

\begin{figure}[t]
	\centering
		\includegraphics[trim = 0 0 0 0in, clip, width=0.68\textwidth]{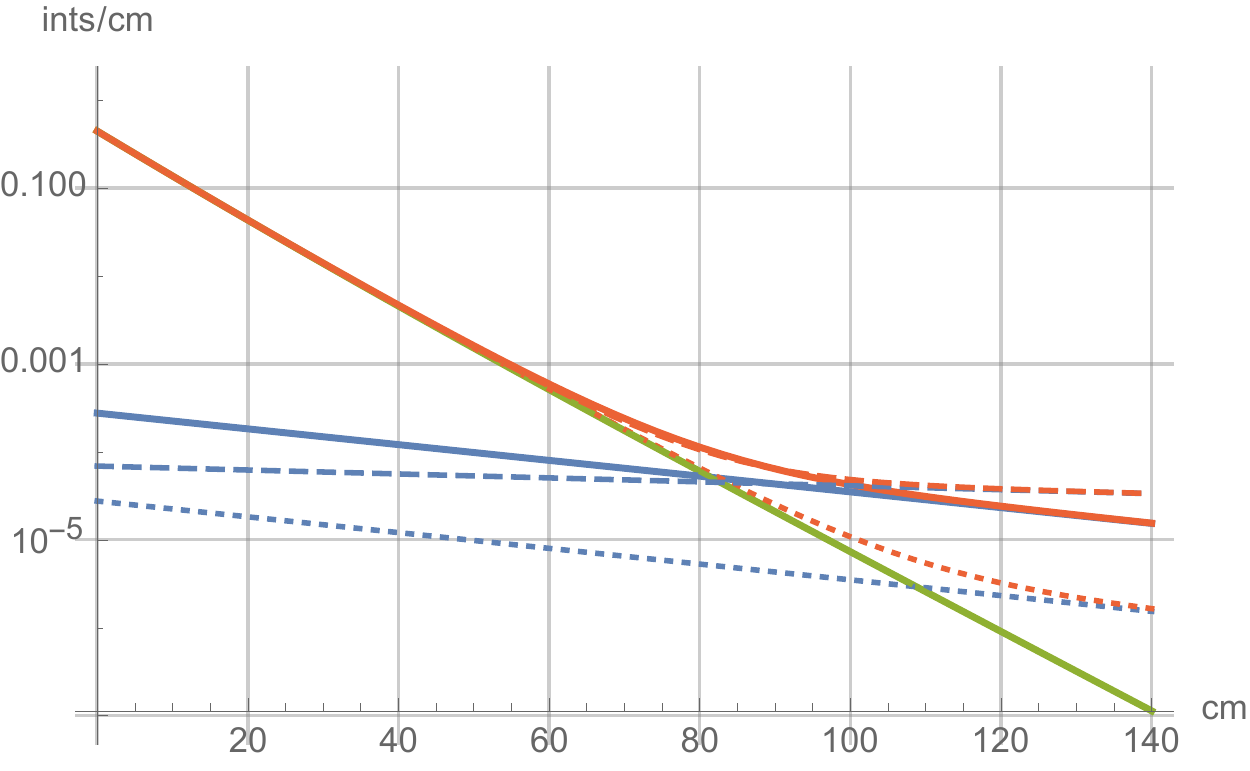}

	\caption{Signal in a detector designed to discover the presence of a new component of neutral  hadrons via a departure from the expected decrease of interactions per unit length.  The horizontal axis is the thickness of material traversed (the brown and black layers in Fig.~\ref{fig:LILN}).  The green curve (obscured by the red curve below $\approx 70$ cm) shows the number of $n$ and $\bar{n}$ scattering and annihilation interactions per cm of Fe or Pb, per produced $n$ or $\bar{n}$, as a function of depth in the absorber in the idealized calculation discussed in the text.  The blue curves show \s\ and \sbar\ interactions per cm of Fe or Pb, for $\lambda_{\rm int, S} = 5  \,\lambda_{\rm int, n}$ (solid) and $\lambda_{\rm int, S} = 20 \, \lambda_{\rm int, n}$ (dashed); in both cases, $\sigma_{\rm prod, S}$ is taken to be  $ 1/300 \, 
	\sigma_{\rm prod, n} $.  The red curves show the total interactions per cm in each case. The dotted lines are for $\lambda_{\rm int, S} = 5 \, \lambda_{\rm int, n}$ and $\sigma_{\rm prod, S} =1/3000 \, \sigma_{\rm prod, n} $. \label{fig:SvsN}}
	\vspace{-0.1in}
  \end{figure}

Under the simplifying approximation that an interacting $n$, $\bar{n}$, \s\ or \sbar\  scatters sufficiently that it will not interact in the detector fiducial region again, the number of interactions per unit length as a function of distance $z$ from the beginning of the absorber-detector is
\be
\label{eq:dNdz}
\frac{dN}{dz} = \frac{1}{\lambda} e^{-z/\lambda}~,
\ee
where $\lambda$ is the interaction length.  Figure~\ref{fig:SvsN} illustrates how the number of interactions per unit length changes as a function of thickness $z$ of the absorber traversed, in the limit that the detector elements and material preceding the absorber-detector device have negligible material; the absorber is taken to be Fe or Pb, with neutron scattering length of 9.7 cm.  (The $n$ and $\bar{n}$ scattering length and the $\bar{ n}$ annihilation length have been inferred from $pp$ and $\bar{p}p$ cross sections at $p_{\rm lab} = 10$ GeV/c, accounting for nuclear shadowing.)	

Figure~\ref{fig:SvsN} shows the number of neutron and anti-neutron interactions per cm of Pb or Fe traversed (green).  The ratio of annihilation and elastic and inelastic scattering cross sections is taken from $pp$ and $\bar{p} p$ scattering at 10 GeV/c.  The blue lines show the number of \s\ and \sbar\ scattering interactions per unit length traversed, taking their interaction length to be five (solid) or twenty (dashed) times that of $n$.  For the solid and dashed curves the \s, \sbar\ production rates are taken to be 1/300 that of neutrons as expected in a central heavy ion collision if $m_S \approx 2 \, m_p$, based on the measured ratio of $p:d$ and $\bar{p}:\bar{d}$ in the central region and assuming particle production can be described as thermal~\cite{Andronic+HIC_Nature18}.  The dotted curves assume a factor-10 lower \s, \sbar\ production rate and $\lambda_{\rm int, S} = 5 \,\lambda_{\rm int, n}$.  

The key point of Fig.~\ref{fig:SvsN} is that only about one meter of absorber is required to observe a deviation from the neutron and antineutron-dominated exponential decrease in interaction rate, even if the \s\ interaction length is large.   Integrating Eq.~\ref{eq:dNdz} yields the total number of $n,\bar{n}$ interactions in any given absorber thickness range, per $n,\bar{n}$ produced in the primary collision and entering the detector.  For instance, in the absorber thickness range $80<z<120\,$cm, the fraction of $n,\bar{n}$'s interacting in that distance range is 0.00057.  For a central heavy ion collision where there are 30 produced n's per unit rapidity, this translates to 0.017 $n,\bar{n}$ interactions per collision per unit rapidity observed by the detector.  For an illustrative choice $\lambda_{\rm int, S} = 5 \, \lambda_{\rm int, n}$ and $\sigma_{\rm prod, S} =1/300 \, \sigma_{\rm prod, n} $, the same range of absorber thickness  would contribute 0.043  \ssbar\ scattering interactions.    With say $10^{10}$ collisions and a geometric acceptance $f \equiv \Delta \phi \, \Delta Y$ this simple estimate gives  $\sim \!1.7\, f \times 10^8$ interactions in the absence of \s\ and \sbar\ versus $\sim \!4.3 \, f \times 10^8$ interactions with \s\ and \sbar, in the given circumstances.  Even though the estimate here is simplistic, e.g., assuming a particle can only scatter once before it disappears, it shows that statistics is not the likely limiting factor of such an approach since there will be plenty of events.  Rather, the sensitivity will be governed by the ability to eliminate background and accurately determine the distance-dependence of $n,\bar{n}$  or \ssbar\ interactions in the absorber.  

As to where such a long-interaction-length neutral detector should be placed, both the fixed-target and the collider setups have advantages and drawbacks.  When the absorber-detector is in the central region in the center-of-mass of a  collision, most $n, \, \bar{n}$, \s\ and \sbar's have energies of a few GeV or less, making their interactions harder to discriminate from background such as decays and cosmic ray secondaries.  On the other hand, in a high energy fixed-target setup, particles such as $K_L$'s with high Lorentz factors could add an additional source of interactions, complicating the simple picture in Fig.~\ref{fig:SvsN}.  Installing the long-interaction-length neutral detector behind an LHC detector or using an existing detector component to achieve the same function, would be valuable for tagging central heavy ion collisions and reducing background.

\begin{figure}[t]
	\centering
		\includegraphics[trim = 0 0 0 0in, clip, width=0.68\textwidth]{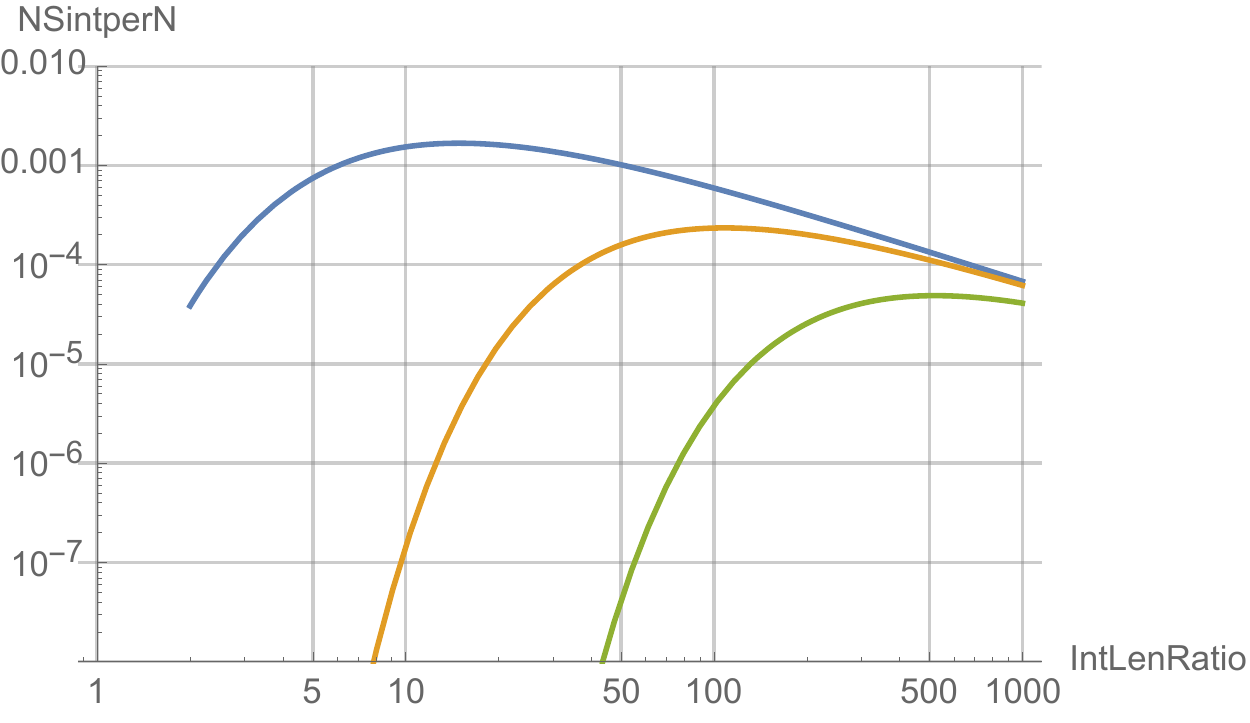}

	\caption{Number of \s\ and \sbar\ interactions in 1m of Fe, per produced neutron in the central region, after 1m (blue), 10m (gold) and 50m (green) of Fe absorber, as a function of the ratio of interaction length of \s\ to that of n, taking $\sigma_{\rm prod, S} =1/300 \, \sigma_{\rm prod, n} $.  
	 \label{fig:IntLenR}}
	\vspace{-0.2in}
  \end{figure}

A third setup can be superior, especially if the \s-nucleon scattering cross section is small.  Namely, put the detector behind more shielding to remove all of the known particles and simply search for events consistent with being a hadronic scattering initiated by an \s\ or \sbar\ coming from the interaction point.  Figure~\ref{fig:IntLenR} shows the number of \s\ and \sbar\ scattering events per neutron produced in the given direction in a 1m Fe absorber-detector such as illustrated in Fig.~\ref{fig:LILN}, as a function of the \s\ scattering length relative to the neutron scattering length, for different amounts of shielding and taking \s\ production to be 1/300 that of $n$.  The figure shows that if  $\sigma_{S-\rm{Fe}} << \sigma_{n-\rm{Fe}} $ so the interaction length ratio is large, there is only a factor-few reduction in event rate going from a shielding thickness of 1m to 50m, according to this simplified estimation framework.  This insensitivity to interaction length ratio can give useful flexibility in locating a suitable site, moreover being able to locate behind a great deal of shielding can be useful for reducing background from the primary interaction products.    

In a detected \s\ or \sbar\ collision the momentum $|\mathbf{p}_S|$ of the incident \s\ is unknown.  However in the central region the spectrum as a function of transverse energy ($E_T \equiv \sqrt{p_T^2 + M^2}$) is similar for all hadrons.  It peaks below a few GeV, so one can reasonably expect the same to be true for \s\ and \sbar.   In the \s-nucleus scattering event, the final \s's 3-momentum is also unknown.  The accompanying secondary hadrons should have a similar distribution as for neutron-initiated events but without the forward diffractive component.  The total energy of the secondaries gives a lower bound on the energy of the incident \s, from which one can infer a lower bound on $v/c$ for a given assumed $m_S$.   One can also model the energy of produced particles as a function of the initial energy of the \s\ and its mass.  With a decent estimate of $v/c$ of the initial \s\ or \sbar, timing the scattering event with respect to bunch-crossings could be useful to reduce background from cosmic ray scattering, and potentially constrain $m_S$.

Practical considerations of cost, as well as physical and engineering constraints, are highly relevant.  We note that the milliQan detector~\cite{milliQan21}, being installed for LHC Run 3  behind 17m of
rock outside the CMS detector, may have sensitivity to \ssbar's produced in the primary collision. The rock would eliminate
the neutron and other remnants of the primary interaction almost entirely, leaving only the remaining \s\ and \sbar\ as neutrals to initiate 
interactions.  These \s\ and \sbar\ initiated vertices could possibly be discriminated from other
backgrounds.  
(Note that the FASER experiment~\cite{FASER21} searched for an anomalous component of neutral penetrating particles which {\it decay,} whereas the sexaquark search needs to trigger on neutral penetrating particles which {\it interact}.)

For milliQan, the detector subtends a fraction 0.0053 of azimuth and the pseudorapidity acceptance is $\pm 0.0167$, giving $f = 8.5 \times 10^{-5}$.  Taking $dN_{n}/dY = 0.17$, we obtain the number of \s\ and \sbar's entering the solid angle of the detector and interacting in 1m Fe-equivalent by multiplying the appropriate curve in Fig.~\ref{fig:IntLenR} by $3 \times 10^{-5}$ times the total number of interactions.   With $10^{16}$ inelastic interactions calculated previously for 150 fb$^{-1}$ and using the gold curve to approximate the milliQan shielding over the range of interaction length ratio shown, the estimated number of events in the milliQan detector is $(3\times 10^4 \,\, {\rm to} \,\,  9 \times 10^7) ( N_{\rm int, mQ}/N_{\rm int, 1 m Fe})$, where the last factor is the ratio of the number of interaction lengths in the milliQan detector to that in the 1m Fe used in Fig.~\ref{fig:IntLenR}.  
An attractive feature of the milliQan setup is that all inelastic interactions would contribute, not just those which trigger the CMS detector.
With modern, very high precision timing and a physically long detector, time of flight could be another tool to discriminate an \ssbar\ component from background.   

\subsection{\sbar\ annihilation in an LHC tracker}
\label{ss:CMS}
Anti-sexaquarks, \sbar, should be produced in high energy $pp$ and heavy ion collisions at the LHC, at a rate of perhaps $10^{-2.5} - 10^{-5}$ relative to anti-neutrons, according to the estimates earlier in this paper.  In a detector such as CMS, ATLAS, ALICE or LHCb, such \sbar's can annihilate with nucleons in the material of the beam-pipe or tracker, to produce a very distinctive final state in which for instance a $\bar{\Lambda}$ and a K emerge from the thin material layer with no visible incoming particle.   

Unfortunately, despite the total number of \sbar's produced being potentially very large thanks to the high integrated luminosity of the LHC and the fact that minimum-bias events are equally good for this study, the expected rate of useful events is small.  If the lab energy of the \sbar\ is small enough for the annihilation event to not produce too many final particles to reconstruct, the annihilation cross section, which goes as $\sim \tilde{g}_{\rm eff}^2$, may be very small.  If the \sbar\  energy is large and many particles are produced, possibly evading the full $\sim \tilde{g}^2$  suppression, then identifying the characteristic B  =  -1 and S = +2 signature of the final state is difficult.   Moreover, reconstruction of the resulting low-momentum tracks with unusual displacement from the interaction vertex proves very difficult, leading to a significant efficiency loss~\cite{thesisjarne}.   See Appendix~\ref{app:SbarAnnih} and \cite{fS17} for more detailed discussion.

Alternative to attempting to reconstruct the final state from the annihilation of the $\bar{S}$ on detector material, one may also attempt to establish an anomalous B=+1 and S=-2 characteristic in the other particles produced in the collision, thus evading the small $\bar{S}$ annihilation cross section. This approach is however hampered by too large backgrounds to be viable.\footnote{S.~Lowette, private communication.}

\section{Summary}   
\label{sec:sum}

An undiscovered stable or extremely long-lived sexaquark composed of $uuddss$ with B=2, S=-2, Q=0 and spin 0 is compatible with our present understanding of QCD. Calculating the mass of the \s\ with realistic quark masses is beyond current capabilities of lattice QCD, and phenomenological models differ widely in their predictions.  Thus experiment must be the arbiter of the existence of a stable sexaquark.


 Experiments rule out an \s\ which is short-lived, but the possibility of an \s\ which is effectively stable has yet to be adequately explored.  The search for a stable sexaquark is challenging because the production rate of \s's is very small in low energy hadronic collisions where quantum number conservation can be readily tracked, due to the short-distance repulsion of baryons which suppresses conversion between two baryons and an \s.  In high energy collisions where the production rate is higher, the \s\ is difficult to identify due to its similarity to the vastly more copious neutrons, moreover baryon number and strangeness accounting are futile due to missing and unidentified particles.  


Strategies to overcome these problems and discover a stable \s\ have been discussed here.  Three are especially promising:
\\
$~~~(1)$  In final states of $\Upsilon$ decay: \\
$~~~~~~~~~(a)$ Search for an excess of events with B=$\pm 2$, S = $\mp2$. \\
$~~~~~~~(b)$ Reconstruct the missing mass in events with $ {\Lambda} {\Lambda}$ or $\bar{\Lambda} \bar{\Lambda}$ and no other baryons or strange particles.  If there is a population of events with an unseen \s\ or \sbar, the missing mass spectrum will have  a characteristic shape which for a perfect, hermetic detector would peak near $m_S$.  \\
$~~~(2)$  Search for $  \gamma \, p \rightarrow S \, \bar{\Lambda} \, K^+ $ (+ possible pions), exploiting the very high luminosity and suitable $E_{\rm cm}$ of GlueX at J-Lab. \\
$~~~(3)$ Perform a search for new, long-interaction-length neutral particles produced in high energy collisions.   Examples of possible detector setups and estimates of rates have been given.  This should be a powerful approach if background can be adequately suppressed.  

The search for a sexaquark -- or more generally any new, long-interaction-length neutral hadron -- is further motivated by our recent observation~\cite{fg-2_22} that production of a yet-unknown neutral, long-lived hadron can resolve the muon g-2 anomaly~\cite{g-2PRL21} by increasing the Hadronic Vacuum Polarization contribution to g-2 relative to the value presently inferred from $e^+e^-$ experiments~\cite{Aoyama+20}.  At the same time, this would reconcile the HVP value measured on the lattice~\cite{BMWHVP21} with observations.  The approach to discovering a hitherto-overlooked long-interaction-length neutral hadron proposed in Sec.~\ref{ss:LIL} directly addresses searching for any such particle, not only the \s.

The possibility that a stable sexaquark may comprise all or part of the dark matter of the Universe adds further urgency to investigating its existence in the laboratory.   

\bigskip

\acknowledgments
I wish to thank my students Z. Wang and X. Xu for collaboration on related aspects of the sexaquark;  
B. Echenard, A. Haas, I. Jaegle, S. Lowette, R. Mussa and S. Olsen 
for information regarding the capabilities of various detectors and discussions and suggestions of search strategies; T. Pierog for implementation of \ssbar\ production in EPOS;  
S. Blaschke, T. Hatsuda, M. Karliner, S. Reddy and T. Schaefer for discussions, and P. Koppenburg for a valuable comment on the manuscript.   
This research was supported by National Science Foundation Grant Nos.~PHY-1212538, PHY-2013199 and the Simons Foundation;  the hospitality of the 
Aspen Center for Physics, supported by NSF-PHY-1607611, is also gratefully acknowledged.  

\def\apj{Astrophys.\ J.}
\def\nat{Nature}
\def\apjl{Astrophys.\ J. Lett.}
\def\apj{Astrophys.\ J.}
\def\aap{Astron.\ Astrophys.}
\def\prd{Phys. Rev. D}
\def\physrep{Phys.\ Rep.}
\def\mnras{Month. Not. RAS }
\def\araa{Annual Rev. Astron. \& Astrophys.}
\def\aapr{Astron. \& Astrophys. Rev.}
\def\aj{Astronom. J.}
\def\jcap{JCAP}

\bigskip

\appendix

\section{Statistical model for inclusive \s\ or \sbar\ production in $\Upsilon$ decay}
\label{app:UpsStat}
The inclusive branching fraction for \s/\sbar\ production in $\Upsilon$ decay was estimated using a statistical model in \cite{fS17} and is reviewed here for completeness.  $\Upsilon$ decays below open-bottom, i.e., $\Upsilon(1S)$, $\Upsilon(2S)$ and $\Upsilon(3S)$ decays, go through 3-gluons.  To leading order each gluon converts to $q$ and $\bar{q}$'s of  $\approx 2$ GeV, which produce mini-jets or form hadrons through string-fragmentation.   Production of the minimum 6$q+6 \bar{q}$ with 6 $q$'s or $\bar{q}$'s having relatively similar momenta needed to produce an \s\ or \sbar\, requires creating 3 more gluons, at a penalty factor of $\alpha_s^3$ in the amplitude.  Here $\alpha_s \gtrsim \alpha_s(\Upsilon) \approx 0.2$, and may be $\mathcal{O}(1)$ because large momentum transfer is not required, so we adopt the geometric mean.    

The next requirement is for 6 $q$ or $\bar{q}$'s to be nearest neighbors in space, within a distance scale $<r_S$.  Statistically, the penalty for having exclusively $q$'s or $\bar{q}$'s within a nearest-neighbor grouping is $\left( \frac{1}{2} \right)^5$.  As a zeroth approximation, no penalty is included for such a grouping of $q$ or $\bar{q}$'s to have an appropriate spatial wavefunction to be an \s\ or \sbar, on the grounds that the $q$ and $\bar{q}$'s originate in a region of size $\approx \frac{1}{10\rm GeV} = 0.02$ fm then expand, so at some point they will be in the relevant volume to form an S. 

Finally, to form an \s\ or \sbar, the 6 $q$'s or $\bar{q}$'s must have the total flavor-spin-color quantum numbers to be an \s\ or \sbar.   Without loss of generality consider the 6$q$ case.  \s\ belongs to the (1,1,1) representation of $SU(3)_c \times SU(3)_f \times SU(2)_s$.  With a spatially symmetric wavefunction, Fermi statistics implies it is in the totally antisymmetric 6-quark representation of $SU(18)$.   In a statistical approximation that the $q$'s produced by the gluons randomly populate all possible color, flavor and spin states, the fraction of cases contributing to the antisymmetric representation is $18\times17\times16\times15\times14\times13/(6\times5\times4\times3\times2)/18^6 =  5.46 \times 10^{-4}$.  Dividing by $2^5$, multiplying by $(\alpha_s^2 \approx 0.2)^3$ and by 2 for the sum of \s\ and \sbar\, results in the estimated branching fraction of Upsilon to states containing an \s\ or \sbar:  $\mathcal{F} \approx 2.7 \times 10^{-7}$.  

A more detailed model calculation could give a larger estimate, since color correlations among the gluons and quarks likely enhance the amplitude for states corresponding to attractive QCD channels, while the simplification of ignoring the spatial structure could lead to a suppression.  Combining these effects with the crude nature of the estimated dependence on $\alpha_s$,  means improvements can take this naive estimate in either direction.  

\section{Exclusive $S\bar{\Lambda}\bar{\Lambda} + c.c.$ branching fraction in $\Upsilon$ decay}
\label{app:UpsExc}
BABAR studied the missing mass in $\Upsilon(2S,3S) \rightarrow  \bar{\Lambda} \, \bar{\Lambda} + single~invisible~particle$ and its charge conjugate, and placed an upper limit on the branching fraction for $\Upsilon(2S,3S)  \rightarrow S \, \bar{\Lambda} \, \bar{\Lambda} + c.c.$.  In this section we take two approaches to estimating theoretically what branching fraction can be expected.  First, we use measured  exclusive branching fractions for $J/\Psi$ and $\Upsilon(1S,2S,3S)$ decays that can provide general guidance for the penalty of demanding a low-particle-number exclusive final state rather than fully inclusive \s\ production.  Second, we start with the measured branching fraction for the exclusive process $\Upsilon_{2,3S} \rightarrow \phi K^+ K^-$ and modify it to to adapt it for $\Upsilon(2S,3S)  \rightarrow S \, \bar{\Lambda} \, \bar{\Lambda}$.

\subsection{Guidance from observed exclusive final states}
General properties of $J/\Psi$ and $\Upsilon$ states are summarized in Table I.  Relevant data from the PDG bearing on  the exclusive penalty in their decays is given in Tables II-IV.  Comments noted below.
\noindent
$J/\Psi$ (Table II):
\begin{itemize}
\item Branching fraction for inclusive baryon-antibaryon production $ \approx 10^{-2}$:  {\it Making a pair of baryons + mesons is 100x harder than just making mesons.}
\item BR($\pi+ \pi^-$)/BR($\pi+\pi^- + X) \approx (1.5 \times 10^{-4})/(0.15) = 10^{-3}$ with $X$ an observed exclusive mode:  {\it Exclusive to inclusive penalty is at least $10^{-3}$.}  
\end{itemize}

\noindent
$\Upsilon_{1S}$ (Table III):
\begin{itemize}
\item Sum over 100 observed exclusive modes accounts for 1.2\% of decays:  {\it Average exclusive  BR $\approx 10^{-4}$. } 
\item There is no observed 2-body decay, with upper limit $4 \times10^{-6}$.  (The most sensitive channels are $\rho \pi, \omega \pi$.)  A few 3-body decays observed, biggest being $\omega \pi^+ \pi^-$ with BR  $4.5 \times10^{-6}$:  {\it Multi-body final states favored over 3-body.  Suppression factor for 3-body at least $\mathcal{O}(10^{-5})$.  }
\item Biggest identified exclusive decay, $\Upsilon_{1S} \rightarrow \pi^+ \pi^-  \,2\pi^0$, has BR $1.3 \times10^{-5}$.
\end{itemize}

\noindent
$\Upsilon_{2S}$, $\Upsilon_{3S}$ (Table IV): 
\begin{itemize}
\item BR($\Upsilon_{2S} \rightarrow \Upsilon_{1S} + \pi \pi $) = 26.5\%.
\item BR($\Upsilon_{3S} \rightarrow \Upsilon_{2S} + X$) = 0.1.
\item Several exclusive 3 body decays seen.  The most significant is $\phi K^+ K^-$ with BR $  1.6 \times 10^{-6}$. 
\item There are only upper limits on $\Upsilon_{2S}$ 2-body decays, except $K^*(892)^0 \bar{K^*_2}(1430)^0+ c.c.$ with BR=  $(1.5 \pm 0.6) \times10^{-6}$.
\end{itemize}

Basically, any individual low-multiplicity exclusive channel in the decay of $ggg$ with invariant mass $\approx$ 10 GeV is highly suppressed.  In $\Upsilon_{1,2,3S}$ decay, respectively 82\%, 59\% and 36\% go through the $ggg$ channel.  From the Tables, the largest 3-body channels detected is consistently VM+2 PS, each having branching fraction $\approx 2 \times 10^{-6}$.  From this, we infer a suppression factor $\approx 10^{-5}$ for a $ggg$ with invariant mass $\approx$ 10 GeV, to go to any 3-body exclusive final state.  Starting from the inclusive branching fraction estimate $2.7 \times 10^{-7}$, gives the estimated branching fraction for the exclusive channels  $\Upsilon_{2,3S} \rightarrow S \bar{\Lambda} \bar{\Lambda} + cc$  of few $ 10^{-12}$. 

\begin{table}[t]
\footnotesize
\begin{tabular}{ | c | c | c|}
\hline
state & mass (MeV) & width (keV)   \\
\hline
\hline
$J/\Psi$ & 3096.9 &  $92.9 \pm 2.8$ \\
$\Upsilon_{1S}$ & 9460.3 &   54\\
$\Upsilon_{2S}$ & 10023.3 &  $32 \pm 2.6$ \\
$\Upsilon_{3S}$ & 10355.2 &  $20.3 \pm 1.85$ \\
\hline\end{tabular}
\caption{{\footnotesize $1^{--}$ Charmonium and Bottomonium states,  from \cite{pdg18} .}}
\end{table}

\begin{table}[t]
\footnotesize
\begin{tabular}{ | c | c | }
\hline
  Final State & branching fraction  \\
\hline
\hline
hadrons & 87.7\% \\
virt. $\gamma$ + hadrons & 13.5\% \\
$ggg$ & 64.1\% \\
$\gamma gg$ & 8.8 \% \\
$\ell^+ \ell^-$ & $\approx 3 \times 2$\% \\
$\pi^+ \pi^- $& $1.47 \times 10^{-4} $\\
$2(\pi^+ \pi^-)\pi^0 $& 4.1\% \\
$3(\pi^+ \pi^-)\pi^0 $& 2.9\% \\
$4(\pi^+ \pi^-)\pi^0 $& 0.9\% \\
$\rho \pi$ & 1.7\%\\
$2(\pi^+ \pi^-\pi^0) $& 1.6\% \\
$\pi^+ \pi^- + X $& 14.6 \%\\
$p \bar{p}$ & $2.1 \times 10^{-3} $\\
$p \bar{p} + \{\pi^0, \pi^+\pi^-,\pi^+\pi^-\pi^0,\eta\}$& $12.5 \times 10^{-3} $\\
$ \omega 2\pi^+ \, 2\pi^-$ & $8.6 \times 10^{-3}$ \\
$\omega \pi^+ \pi^- \pi^0, \, \omega \pi^0 \pi^0$ & $4.0, \,3.4 \times 10^{-3}$ \\
$\pi^+ \pi^- \pi^0, \, \phi K^+ K^- , \, K^+ K_S^0 \pi^-$ & $\approx 2 \times 10^{-6}$   \\
$\pi^+ \pi^-  \,2\pi^0$ & $1.3 \times 10^{-5}$\\
$\Xi^0  \bar{\Xi^0} ,~\Delta(1232)^{++}  \bar{\Delta(1232)}^{--}$  & $1.7,\, 1.1 \times 10^{-3}$\\
\hline\end{tabular}
\caption{{\footnotesize $J/\Psi_{1S}$,  from  \cite{pdg18} }}
\end{table}

\begin{table}[t]
\footnotesize
\begin{tabular}{ | c | c | }
\hline
  Final State & branching fraction  \\
\hline
\hline
$ggg$ & 81.7\% \\
$\gamma gg$ & 2.2 \% \\
$\ell^+ \ell^-$ & $\approx 3 \times 2$\% \\
$\rho \pi,\, \omega \pi$ & $< \approx 3.7 \times 10^{-6}$  \\
$\pi^+ \pi^-, \,K^+ K^- , \, p \bar{p}$ & $< 5 \times 10^{-4}$  \\
$\pi^+ \pi^- \pi^0, \, \phi K^+ K^-  \,( \omega \pi^+ \pi^-)$ & $\approx 2(4.5) \times 10^{-6}$ \\
$\pi^+ \pi^-  \,2\pi^0$ & $1.3 \times10^{-5}$\\
sum 100 exc. modes & 1.2 \% \\
$J/\Psi + X$ & $(5.4 \pm 0.4) \times 10^{-4}$ \\
\hline\end{tabular}
\caption{{\footnotesize Selected $ \Upsilon_{1S}$ hadronic decays,  from  \cite{pdg18}}}
\end{table}

\begin{table}
\footnotesize
\begin{tabular}{ | c | c | }
\hline
  Final State & branching fraction \\
\hline
\hline
$ggg$ & 58.8\% \\
$\gamma gg$ & 8.8 \% \\
$\Upsilon(1S) + \pi \pi $ & 26.5 \% \\
$J/\Psi$+anything & $< 6 \times 10^{-3}$  \\
$\pi^+ \pi^- 2 \pi^0$ & $1.3\pm 0.3  \times 10^{-5}$ \\
$ \phi K^+ K^- $ & $(1.6 ) \pm 0.4) \times 10^{-6}$ \\
$ K^* K^- \pi^+$ & $ (2.3  \pm 0.7) \times 10^{-6}$ \\
$K_S^0 K^+ \pi^-$ + cc  & $1.14 \times 10^{-6}$ \\
\hline\end{tabular}
\caption{{\footnotesize Selected $\Upsilon_{2S}$,  from  \cite{pdg18} }}
\end{table}

\begin{table}
\footnotesize
\begin{tabular}{ | c | c | }
\hline
  Final State & branching fraction  \\
\hline
\hline
$ggg$ & $0.357 \pm 0.026$ \\
$\gamma gg$ & $0.0097 $   \\
$\Upsilon_{2S}$+anything & 0.106  \\
(non-$\Upsilon_{2S}) b\bar{b}$+anything & 0.32  \\
$\ell^+ \ell^-$ & $\approx 3 \times 0.02$  \\
\hline\end{tabular}
\caption{{\footnotesize Selected $\Upsilon_{3S}$,  from  \cite{pdg18}.  No non-leptonic non-$b$-containing final states seen. }}
\end{table}

\begin{table}[b]
\footnotesize
\begin{tabular}{ | c | c | c |}
\hline
  Final State & branching fraction & number events \\
\hline
\hline
$\gamma \, \pi^+ \pi^-$ & $6.3 \pm 1.2 \pm 1.3 \times 10^{-5}$ &\\
$\gamma \, 2\pi^+ 2\pi^-$ & $2.5 \pm 0.7 \pm 0.5 \times 10^{-4}$ & 26 \\
$\gamma \, \pi^+ \pi^- K^+ K^-$ & $2.9 \pm 0.7 \pm 0.6 \times 10^{-4}$ & 29 \\
$\gamma \, \pi^+ \pi^- p \bar{p}$ & $1.5 \pm 0.5 \pm 0.3 \times 10^{-4}$ & $7 \pm 6$ \\
$\gamma \, \pi^0 \pi^0$ & $1.7 \times 10^{-5}$ &\\
$\gamma \, K^+ K^-$ & $< 1.14 \times 10^{-5}$ & 90\% CL\\
$\gamma \, p \bar{p}$ & $< 0.6 \times 10^{-5}$ & 90\% CL\\
$\gamma \, 2h^+ 2h^-$ & $7.0 \pm 1.1 \pm 1.0 \times 10^{-4}$ & $80 \pm 12$\\
$\gamma \, 3h^+ 3h^-$ & $5.4 \pm 1.5 \pm 1.3 \times 10^{-4}$ & $80 \pm 12$\\
$\gamma \, 4h^+ 4h^-$ & $7.4 \pm 2.5 \pm 2.3 \times 10^{-4}$ & $80 \pm 12$\\
$\gamma \, 2K^+ 2K^-$ & $1.14 \times 10^{-5}$ & $2 \pm 2$\\
\hline\end{tabular}
\caption{{\footnotesize Radiative $\Upsilon_{1S}$, CLEO measurements from  \cite{pdg18} }}
\end{table}

\subsection{Direct estimate starting from BF($\Upsilon_{2S,3S} \rightarrow \phi K^+ K^-)$ }
We seek to estimate BF($\Upsilon_{2S,3S} \rightarrow S \bar{\Lambda} \bar{\Lambda}) $ starting from the branching fraction for an observed 3-body decay.  At least one pair of final particles in $\Upsilon \rightarrow S \, \bar{\Lambda} \, \bar{\Lambda}$ must have $L=1$ to satisfy angular momentum, parity conservation and Fermi statistics.  This suppresses the rate compared to an $L=0$ final state like $\phi K^+ K^-$, but the 10 GeV CM energy has large phase space, so we ignore it.  

The BF($\Upsilon_{1S,2S,3S} \rightarrow \phi K^+ K^-) \approx 2 \times 10^{-6} $ for all three initial states.   Following the discussion for inclusive \s\ or \sbar\ production, the rate for  $\Upsilon_{2S,3S} \rightarrow ggg \rightarrow S \bar{\Lambda} \bar{\Lambda} $ is suppressed relative to $\Upsilon_{2S,3S} \rightarrow ggg \rightarrow \phi K^+ K^-$ by the factors
\begin{enumerate}
\item $(\alpha_s^2 \approx 0.2)^3 = 0.008 $, to account for the production of 3 additional $q \bar{q}$ pairs beyond the 3 $q \bar{q}$ pairs required to form any non-exotic hadronic final state from $ggg$ (which is in particular sufficient to form $\phi K^+ K^-$).   
\item $(1/2)^5 = 0.03$, the probability of having 6$q$'s or 6$\bar{q}$'s in immediate proximity to make an \s\ or \sbar.  The parallel factor for the $\phi K^+ K^-$ final state is a factor $(1/2)^1 = 0.5$.
\item The color-flavor-spin factor from the requirement that the 6$q$'s or 6$\bar{q}$'s have the quantum numbers of the \s\ or \sbar\, the totally antisymmetric singlet representation of SU(18): $18\times17\times16\times15\times14\times13/(6\times5\times4\times3\times2)/18^6 =  5.46 \times 10^{-4}$.   For the case of the $\phi$, the analogous factor is $(1/3)(3/4)(1/9) =  0.028$, where the factors are respectively for forming a color singlet, spin-1 and specified flavor.
\end{enumerate}
Thus ignoring spatial wave-function suppression we get the estimate
\begin{align*}
{\rm BR}_{\Upsilon_{2,3S} \rightarrow S \bar{\Lambda} \bar{\Lambda}} = &  \, 0.008  \left(\frac{0.03}{0.5} \cdot \frac{5.4 \,10^{-4}}{2.8 \, 10^{-2}} \right){\rm BR}_{\Upsilon_{2,3S} \rightarrow \phi K^+K^-} \\
= &1.25 \times 10^{-11}.
\end{align*}
Whether there is also a penalty for making $p p$ instead of $K^+ K^-$ is unclear.  No such penalization is called for when calculating inclusive \s\ production,  because baryon number conservation requires production of a pair of anti-baryons with \s.   However for the exclusive amplitude the penalization may apply, since there is no unitarity sum over all possible final states.  In that case, there would be an additional suppression $\mathcal{O}(10^{-1})$ as seen in $J/\Psi$ decay and the heuristic 0.1 for a baryon relative to a meson high-multiplicity hadronic final states.  We also dropped the flavor-counting-factor since that is less important.  In sum, this approach leads to the estimate BF($S \bar{\Lambda} \bar{\Lambda}) \approx \mathcal{O}(10^{-11}-10^{-12})$.

Thus two different estimation methods point to a similar branching fraction for the $S \bar{\Lambda} \bar{\Lambda} $ and c.c. exclusive final states, $\mathcal{O}(10^{-11}-10^{-12})$, which is 4 orders of magnitude or more below the current experimental limit set by BABAR  \cite{babarS18}.

\section{Feasibility of using $B-S = \pm 4$ as a diagnostic for \s\ production}
\label{app:B-S=4}

In the ``maximally inclusive" approach, the signature is the imbalance of baryon number B and strangeness S associated with production of an undetected  \s\ or \sbar.  (The probability of production of both an \s\ and an \sbar\ in a single event is small and can be ignored for this purpose.)  The \s\ or \sbar\ being neutral, stable and not interacting in the detector, escapes undetected, with its B and S compensated by other particles in the event.   
With an estimated inclusive branching fraction of $\approx 3 \times 10^{-7}$ in $\Upsilon(1S,2S,3S)$ final states,  and few-body exclusive branching fractions typically a factor $10^4$ or more smaller than their inclusive counterparts (see App. \ref{app:UpsExc}) it is motivated to ask whether searching for an excess of events with $B-S = \pm 4$ could be a possible strategy.  Or, more generally, comparing the relative abundances of events with B-S combinations that are rare but get contributions from events when an \s\ or \sbar\ is produced and escapes, with its quantum numbers balanced in the remaining hadrons.  

A proper study requires detector simulations, but here we assess the feasibility taking a single ``effective efficiency" $e_b$ for identifying a produced baryon or anti-baryon as such and ``effective efficiency" $e_s$ for identifying a produced strangeness+1 or strangeness-1 particle as such.  The other needed inputs are $f_b$, the branching fraction for producing a $B-\bar{B}$ pair in the final state, and $f_s$, the branching fraction for producing a strange-anti-strange pair in the final state, in the absence of \s\ or \sbar\ production.  From Z-decay final states, we adopt for our rough estimate $f_b = 0.05, \, f_s = 0.16$.  We make the further simplifying assumption that the inclusive branching fraction to produce a pair of baryons (in the absence of \s\ or \sbar\ production) is $f_b^2$, and so on.  The branching fraction to produce a single \s\ or \sbar\ is  $\mathcal{F}/2$.  In the following, we take $\mathcal{F}<<f_b < f_s$ and keep higher orders in $f_b$ and $f_s$ such that the dropped terms are less significant than those kept, assuming $e_b, \, e_s > 0.2$.

As a simplifying assumption, we take the efficiency of correctly recognizing a $B=+1$ and $B=-1$ particle to be equal, which is not exactly true, and similarly for strangeness $\pm1$.   For this rough estimate, we assuming identifying baryons and strangeness is independent and uncorrelated, but that is not exact -- for instance $\Lambda, \bar{\Lambda}$'s are particularly easy to ID but in the approximation adopted, the efficiency of counting them would be $e_b \, e_s$, also missing the correlation due to the fact that the geometric penalty should not be counted independently.  In fact, for those reasons, the estimate below may be somewhat pessimistic.  Likewise, studies focussing specifically on correlations in hyperon production (e.g., excess of events with observed $\Lambda, \Lambda$ or $\bar{\Lambda}, \bar{\Lambda}$ compared to expectations from $\Lambda, \bar{\Lambda}$) may prove to be more favorable due to their enhanced ID power, overcoming their more limited statistics.

Using the abbreviation 
\{x$b$ y$\bar{b}$ z$s$ v$\bar{s}$\} to represent the fraction of final states in which there are exactly x identified baryons, y identified anti-baryons, z identified S=+1 particles and v identified S=-1 particles, this simplified treatment enables us to write all of the  \{x$b$ y$\bar{b}$ z$s$ v$\bar{s}$\}'s up to final states with exactly 4 identified B and/or S non-zero particles.  The following gives some examples:                                                                       
\begin{align}
 \{1b\} &= e_b (f _b(1 - e_b) + \mathcal{F} (1 - e_b) (1 - e_s)^2 + 
    2 f_b^2 (1 - e_b)^3) (1 + f_s (1 - e_s)^2) \\
    \{2b\,2\bar{s}\} &= e_b^2 e_s^2 (\mathcal{F}/2 + f_b^2 f_s^2 (1 - e_b)^2 (1 - e_s)^2) \\
   \{ 1b\,1\bar{b}\,1s\,1\bar{s}\} &= e_b^2 e_s^2 (f_b f_s + 4 f_b f_s^2 (1 - e_s)^2 + 4 f_b^2 f_s (1 - e_b)^2) ~.
\end{align}
This treatment does not consider mis-identification which must be taken into account for a complete assessment, but it gives an idea of the possible statistical power.  

The most sensitive combination to find a signal for $\mathcal{F} \neq 0$ is of course $ \{2b\,2\bar{s}\}  = \{2\bar{b}\,2s\} $.  To assign a statistical significance to the measurement of a non-zero value of $\mathcal{F}$ in terms of ``number of $\sigma$'s", we use
\be
n_\sigma(\mathcal{F}) = \frac{( \{2b\,2\bar{s}\}(\mathcal{F}) +  \{2\bar{b}\,2s\}(\mathcal{F}) - (\{2b\,2\bar{s}\}(\mathcal{F}=0) + \{2\bar{b}\,2s\}(\mathcal{F}=0))\sqrt{ N_{\rm tot}}}{\sqrt{(\{2b\,2\bar{s}\}(\mathcal{F}) +  \{2\bar{b}\,2s\}(\mathcal{F}) + \{2b\,2\bar{s}\}(\mathcal{F}=0) + \{2\bar{b}\,2s\}(\mathcal{F}=0))/2}} ~.
\ee
Figure \ref{fig:NsigvsBF} shows the significance by which a given sexaquark branching fraction is distinguished from background according to this simplified analysis, for $N_{\rm tot} =10^9$.  The statistical model estimate for the inclusive branching fraction of \s\ plus \sbar\ is $\mathcal{F}=2.7 \times 10^{-7}$ and the significance scales as $\sqrt{N_{\rm tot}}$.   
The different colored lines give 3 examples of how the significance depends on the efficiencies $e_b$ and $ e_s$.   The Belle-II particle ID efficiency for baryon and kaons is 0.8-0.9 depending on how hard they cut, which depends on what backgrounds they want to suppress, so we consider $e_b = 0.8$ and $0.9$.  Since only charged K's have a definite strangeness, we penalize kaons by a factor-2; hyperons on the other hand have clearly determined strangeness, so we consider $e_s=0.4$ and $0.5$.  Not surprisingly, large $e_b$ is more important than large $e_s$, since baryons are rarer and thus any sexaquark contribution makes a larger relative impact on the baryon abundances. 

A detailed analysis could exploit other combinations such as $ \{2b\,1\bar{s}\}$ which have some sensitivity to $\mathcal{F}$ but with a larger fractional contribution from standard channels, and use cases like $ \{ 1b\,1\bar{b}\,1s\,1\bar{s}\}$  which get no contribution from \s\ or \sbar\ to develop confidence that the systematics of the background are fully understood.  How well the systematics deriving from the matter-antimatter asymmetry of the detector are understood, can be checked as well.  Since the initial $\Upsilon$ has B=S=0 we know the produced particles are symmetric under (B-S)$\rightarrow$-(B-S).  But since  negative baryon number particles have some probability of annihilating in the beam pipe or other material, which raises the apparent baryon number in the final state by one unit if not accounted for, events satisfying  B-S = +4 or +2 could somewhat outnumber those satisfying B-S = -4 or -2, with a corresponding enhancement to B-S = -3 or -1 relative to B-S = +3 or +1.

\section{Selected other experiments}
\label{app:otherexpts}

\subsection{Experiment excluding a long-lived loosely-bound $H$ dibaryon, without decay or mass requirement}
The BNL E888 collaboration performed two different searches for the $H$ dibaryon~\cite{belz+96} and \cite{belz+DiffDissocH96}.  The latter uniquely among existing experiments, did not restrict its search to mass greater than 2 GeV or to decaying $H$ dibaryon.   

In E888, a 24.1 GeV/c proton beam from the Brookhaven National Laboratory AGS, was directed onto a Pt target from which a neutral beam cleaned of photons was produced at 65 mr.  Approximately 18 m down stream, it struck a scintillation counter system -- ``the dissociator" -- which was followed by drift chambers, magnet, trigger counters and Cerenkov counter, optimized for identifying diffractively produced $\Lambda$'s.  A total of $4 \times 10^7$ events were recorded. They placed a limit on the product of cross sections for $H$ production and dissociation:
\be
\frac{d \sigma_H}{d \Omega}|_{\rm 65\, mr} < \left( \frac{N^H_{\Lambda \Lambda} A_{\Lambda K_S} \sigma^c_{\Lambda K_S}}{N^c_{\Lambda K_S} A_{\Lambda \Lambda} \sigma_{\Lambda \Lambda}}\right) ~ \frac{d \sigma_n}{d \Omega}|_{\rm 65\, mr}~,
\ee
where $N^c_{\Lambda K_S}$ is the number of coherently produced $\Lambda^0 K^0_S$, $A_{\Lambda \Lambda}$ and $A_{\Lambda K_S} $ are the acceptances and efficiencies for $H + A \rightarrow \Lambda^0 \Lambda^0 A \rightarrow p \pi^- p \pi^- A$ and $ n + A \rightarrow \Lambda^0 K^0_S X \rightarrow p \pi^- \pi^+ \pi^- X$ respectively, where diffractively dissociated final states are characterized by the distinctively forward-peaked distribution, and $\sigma_{\Lambda \Lambda}$ and $\sigma^c_{\Lambda K_S}$ are the respective diffractive dissociation cross sections.   Using their acceptance estimates, upper limits on $N^H_{\Lambda \Lambda}$ taking into account estimates of backgrounds, and knowledge of  $\frac{d \sigma_n}{d \Omega}$, they found for 24.1 GeV/c p-Pt collisions at 65 mr:
\be
\frac{d \sigma_H}{d \Omega}|_{\rm 65\, mr} ~ \frac{\sigma_{\Lambda \Lambda}}{0.5 {\rm mb}} < 2.3\times 10^{-4}   \frac{d \sigma_n}{d \Omega}|_{\rm 65\, mr}~\frac{\sigma^c_{\Lambda K_S}}{5.9 \mu {\rm b}} ~.
\ee

The conventional H-dibaryon scenario, in which the H is a relatively loosely bound state of two $\Lambda$'s, is constrained by these limits.  However this limit is far too weak to be constraining for the tightly bound \s\ scenario.  The diffractive dissociation cross section has a wavefunction overlap penalty of \gtilde$^2$; as discussed in the main text, this is $\lesssim 10^{-11}$.  
For a review of eight other H-dibaryon searches, see \cite{chrienHBNL98}.   

\subsection{Doubly-strange hypernucleus decay}
\label{app:hypernuc}
By now, the study of the masses of various doubly-strange hypernuclei has become an important research area in nuclear physics, providing a new window on $\Lambda-\Lambda$ and $\Xi-N$ interactions.   The KEK group has developed an extensive dataset of events recorded in an emulsion with a sequence of reconstructed interactions providing strong constraints on the kinematics and interpretation. One famous example is the ``Nagara event"~\cite{takahashi+hypernucKEK01,Ahn+DoubLamHypernuc13}; it has multiple interpretations, but all involve existence of a double-$\Lambda$ nucleus which decays to a single-$\Lambda$ nucleus and nucleons.  Other events confirm the production of hypernuclei with two $\Lambda$'s.  Such processes demonstrate that the time-scale for conversion of a $\Lambda \Lambda$ system into an \s\ is longer than for the weak decay of one of the $\Lambda$'s.  This excludes the existence of a loosely bound H-dibaryon whose overlap with a di-$\Lambda$ state would be large,  but is compatible with expectations for a sexaquark as discussed in Sec.~\ref{sec:gtilde}.  Taking the formation time of an \s\ to be  $\gtrsim 10^{-10}$s, places an observational bound \gtilde\ $\lesssim 10^{-5}$.  As seen in Fig.~\ref{fig:gtilde} this is consistent with expectations for a sexaquark.

\subsection{\sbar\ annihilation in the LHC beam-pipe or detector} 
\label{app:SbarAnnih}

A possible strategy~\cite{fS17} is to take advantage of the tremendous luminosity of the LHC, and look for characteristic decay chains after \sbar\ annihilation in the beam-pipe or detector, e.g., 

\begin{eqnarray}
\label{Sbarannih}
\bar{S}+N &\rightarrow& \bar{\Xi}^{+,0} + X, \,{\rm with}\,\, \bar{\Xi}^{+,0} \rightarrow \bar{\Lambda} \pi^{+,0} \,\&\, \bar{\Lambda} \rightarrow \bar{p} \pi^+~  \nonumber \\{\rm or}~
\bar{S}+N &\rightarrow& \bar{\Lambda} + K^{+,0} + X.
\end{eqnarray}

$\bar{S}$'s should have a similar transverse-energy distribution as other hadrons, i.e., $< \! p_t  \!>\lesssim \mathcal{O}(1\,$GeV).  The anti-baryon produced in $\bar{S}$ annihilation in a central tracker will commonly be a $\bar{\Xi}^{+,0}$ with $\gamma  \!\sim \! 1$ which decays in $\mathcal{O}(5\,$cm) to $\bar{\Lambda} \pi^{+,0}$, followed 64\% of the time by $\bar{\Lambda} \rightarrow \bar{p} \pi^+$ in about 8 cm.  Observing such distinctive production and decay chains provides unambiguous evidence for a B=-2, S= +2 neutral particle that initiated the annihilation interaction in the beam pipe or detector.  Complementing $\bar{\Xi}$ production are $\bar{S} n \rightarrow \bar{\Lambda} K^0_S$, where the double ``$V$"s locate the annihilation vertex and $\bar{\Lambda}$ identification proves the initiating particle has $B=-2$, and $\bar{S} p \rightarrow \bar{\Lambda} K^+$, where the $B=-2$ and $ S=+2$ of the initial state is unambiguous.

There are $\approx \! 30$ charged particles with pseudo-rapidity $|\eta | < 2.4$, for 7 TeV LHC  p-p collisions \cite{CMSchgpartmult10}, so the number of \sbar's produced with pseudo-rapidity $|\eta | < 2.4$ in a dataset with $N_{11}10^{11}$ recorded interactions is  
$ 
N_{\bar{S}} \approx  3 f^{\rm prod}_{-4} \, N_{11} \, 10^{8},
$
where $f^{\rm prod}_{-4} 10^{-4}$ is the \sbar\ production rate relative to all charged particles in the given rapidity range.  The material budget for the CMS tracker and beam pipe ranges from 0.12-0.55 hadronic interaction lengths in this $\eta$ range; take 0.33 for an estimate.  Write the \sbar\ annihilation cross section as $\sigma_{\bar{S}N} \equiv f^{\rm annih}_{-6} 10^{-6} \sigma_{NN} $, acknowledging that the annihilation reaction is at relatively low energy so the penalty of the overlap suppression is more severe than that for production in the original high energy collision.  Finally, take the fraction of annihilation final states containing $\bar{\Xi},\bar{\Lambda}$ to be $f_{\bar{\Xi},\bar{\Lambda}}$, where we expect $f_{\bar{\Xi},\bar{\Lambda}} \sim \mathcal{O}$(1) for low $\sqrt{s}$ annihilation.  Thus the number of potentially reconstructable annihilation+decay chains is 
\be
 N_{\bar{\Xi},\bar{\Lambda}} \, = \, f^{\rm prod}_{-4} \, f^{\rm annih}_{-6} \, f_{\bar{\Xi},\bar{\Lambda}} \, N_{11} ~~ 10^{2}.
\ee
 
In some subset of events, all the final particles of the annihilation will be identified and their 3-momenta adequately measured.  With these events, the 3-momentum and kinetic energy of the \sbar\ can be deduced from energy-momentum conservation, modulo the nuclear Fermi-momentum of the nucleon on which the \sbar\ annihilates.  In principle, this could enable the mass of the \sbar\ to be measured.  However reconstruction of the events proved difficult in practice~\cite{thesisjarne}.  Production of \s\ and \sbar\ has been implemented in the microcanonical fireball contribution to final states in the event generator EPOS-LHC, which will facilitate detector studies for LHC and for $\Upsilon$ decay~\cite{pierog+22}.


\bibliography{nucphys,QCD,H,f,dm,SDM,pheno,anoms}
\bibliographystyle{JHEP}

\end{document}